\documentclass[sn]{sn}

\usepackage{graphicx}%
\usepackage{multirow}%
\usepackage{amsmath,amssymb,amsfonts}%
\usepackage{amsthm}%
\usepackage{mathrsfs}%
\usepackage[title]{appendix}%
\usepackage{xcolor}%
\usepackage{textcomp}%
\usepackage{manyfoot}%
\usepackage{booktabs}%
\usepackage{algorithm}%
\usepackage{algorithmicx}%
\usepackage{algpseudocode}%
\usepackage{listings}%

\theoremstyle{thmstyleone}%
\theoremstyle{thmstyletwo}%

\theoremstyle{thmstylethree}%

\begin{document}

\title[Article Title]{Estimating the epidemic threshold under individual vaccination behaviour and adaptive social connections: A game-theoretic complex network model}


\author[1]{\fnm{Viney} \sur{Kumar}}

\author[2]{\fnm{Chris T} \sur{Bauch}}

\author*[1]{\fnm{Samit} \sur{Bhattacharyya}}\email{samit.b@snu.edu.in}

\affil[1]{\orgdiv{Department of Mathematics}, \orgname{School of Natural Science, Shiv Nadar Institution of Eminence}, \orgaddress{\street{NH-91}, \city{Greater Noida}, \postcode{201314}, \state{Uttar Pradesh}, \country{India}}}

\affil[2]{\orgdiv{Department of Applied Mathematics}, \orgname{University of Waterloo}, \orgaddress{\street{200 University Avenue West}, \city{Waterloo}, \postcode{N2L 3G1}, \state{ON}, \country{Canada}}}


\abstract{Information dissemination intricately intertwines with the dynamics of infectious diseases in the contemporary interconnected world. Recognizing the critical role of public awareness, individual vaccination choices appear to be an essential factor in collective efforts against emerging health threats. This study aims to characterize disease transmission dynamics under evolving social connections, information sharing, and individual vaccination decisions. To address this important problem, we present an integrated behaviour-prevalence model on an adaptive multiplex network. While the physical layer (layer-II) focuses on disease transmission under vaccination, the virtual layer (layer-I), representing individuals' social contacts, is adaptive and deals with information dissemination, resulting in the dynamics of vaccination choice in a socially influenced environment. Utilizing the microscopic Markov Chain Method (MMCM), we derive analytical expressions of the epidemic threshold for populations with different levels of perceived vaccine risk. It indicates that the adaptive nature of social contacts contributes to the higher epidemic threshold compared to non-adaptive scenarios, and numerical simulations also support that. The network topology, such as the power-law exponent of a scale-free network, also significantly influences the spreading of infections in the network population. We also observe that vaccine uptake increases proportionately with the number of individuals with a higher perceived infection risk or a higher sensitivity of an individual to their non-vaccinated neighbours. As a result, our findings provide insights for public health officials in developing vaccination programs in light of the evolution of social connections, information dissemination, and vaccination choice in the digital era.}


\maketitle

\section{Introduction}\label{sec1}

In the recent digital era, social connections and community-level awareness play a major role in the success of any policy implementation, such as vaccinations or social distancing to control infectious diseases \cite{argyris2023social}. In voluntary vaccination programs, while access to vaccines and individual choices are essential for the effectiveness of the program \cite{rodrigues2020impact,deroo2020planning}, the information and rumors over social platforms play crucial roles in shaping the dynamics of vaccine uptake \cite{zhang2022social,chen2022spread}. In the face of an outbreak, each community member engages in a cognitive evaluation, formulating a unique assessment of the likelihood of contracting the virus and the perceived effectiveness of vaccination depending on information obtained from various social media platforms \cite{tong2021cognitive}. Personal experiences, complex information distribution channels, and societal influences are just a few of the variables that have an impact on this multifaceted decision-making framework \cite{dube2018underlying,wiedermann2023unravelling}. For instance, a study conducted in the U.S. revealed that approximately 30 \% increase in the proportion of individuals willing to vaccinate as the perceived infection risk rises due to information from various social platforms \cite{baumgaertner2020risk}. On the other hand, social connections can also evolve rapidly, particularly during global or community-level emergencies, such as a disease outbreak like COVID-19 \cite{folk2020did,alizadeh2023impacts}. While similar levels of risk perception can strengthen virtual bonds among community members by fostering communal understanding and solidarity \cite{gu2021assessing}, disproportionate risk perceptions can cause individuals to avoid existing connections and actively seek new ones \cite{bascunan2022convivial}. Thus, evolution of virtual connections is a common and natural phenomenon that facilitates the exchange of diverse perspectives and new information on vaccination. This however, plays a major role in influencing individual vaccination decisions \cite{oestereich2023optimal}.\\

Several mathematical models have been explored to analyze how individual awareness of vaccines affects their actions during a disease outbreak \cite{d2020interplay,shanta2020impact,buonomo2019optimal,alam2019game}. d’Onofrio et al.\cite{d2020interplay} established a general framework of behaviour-dependent linear and nonlinear forces of infection (FOI) that accounts for lags and effects due to \textit{epidemic memory}. Sharmin et al. \cite{shanta2020impact} developed a SIR-type generic model that includes \textit{media literacy} as a moderator to socially isolate sick and vulnerable populations during an outbreak. Buonomo et al. \cite{buonomo2019optimal} statistically examined how human behaviour, seasonality, and latency rate impact the containment and spread of infection. Several mathematical and computational models have been devised later that focused on population heterogeneity as the key element \cite{alam2019game,yin2022impact,kuga2018impact,wang2020vaccination,wang2020roles}. Yin et al. \cite{yin2022impact} developed a three-layer coupled network model to study the co-evolution of negative vaccine-related information, vaccination behaviour, and epidemic spread. Juan Wang et al. \cite{wang2020roles} developed a two-layered network strategy for family networks that accounts for both conformity-motivated updates and myopic best response-motivated updates. Xuelong Li et al. \cite{wang2020vaccination} created a two-layered multiplex network model that mimics human nature while accounting for the SIR epidemic process and vaccination decision-making. Additionally, a handful of mathematical models have been developed to study the impact of adaptive networks in a single-layered network \cite{gross2006epidemic, marceau2010adaptive}. Their work investigated epidemic dynamics on an adaptive network, where susceptibles could avoid contact with the infected by rewiring their network connections. Recently, Xiong et al. \cite{xiong2023epidemic} developed a UAU-SIS model to simulate the diffusion of awareness and epidemic spreading on temporal multiplex networks. 

However, the theoretical framework of infectious disease dynamics, which explores human choice in vaccination while simultaneously sharing opinions under the adaptive nature of social connections, is very rare. Also, there is rarely any modelling study that characterizes the  emergence of multiple epidemic thresholds and its characterization due to variations in vaccinations risk perceptions among individuals in community.\\ 

In this study, we have developed a framework integrating adaptive multiplex networks and game theory to tackle these research questions. Our adaptive game-theoretic behaviour-prevalence multiplex network model, which combines the intricacies of information dissemination, decision-making, and the dynamics of disease transmission across a diverse population, can shed light on designing intervention strategies or policy-making in controlling disease. We have employed  Microscopic Markov chain Method (MMCM) \cite{gomez2010discrete}, to compute the epidemic threshold by levels of perceived vaccine risk in infected states. We observed that the adaptive nature of social connections has disproportionate impact on different epidemic thresholds with different risk perceptions. Our results have shown that the topology of networks, such as an increase in power-law exponents in scale-free networks, increases the cumulative infection. It is also seen from our findings that the vaccine uptake also increases with a high number of unvaccinated neighbours with high-risk perceptions, which increases the vaccine uptake in the community. So taken together, this research underscores the significance of the adaptive nature of network dynamics and decision-making processes concerning perceived vaccine and infection costs during a disease outbreak. These findings have potential insights for organizing vaccination campaigns in the community.\\

The article is framed as follows: Section 2 describes the components of the model and their integration; Section 3 elucidates the computation of the epidemic thresholds using MMCM, computational simulation algorithm, and numerical simulations of the epidemic thresholds, network metrics, and diverse epidemiological parameters. Finally, in Section 4, we draw the conclusions and limitations.

\section{Model framework}
\begin{figure}[h]
\centering
\includegraphics[scale=.35]{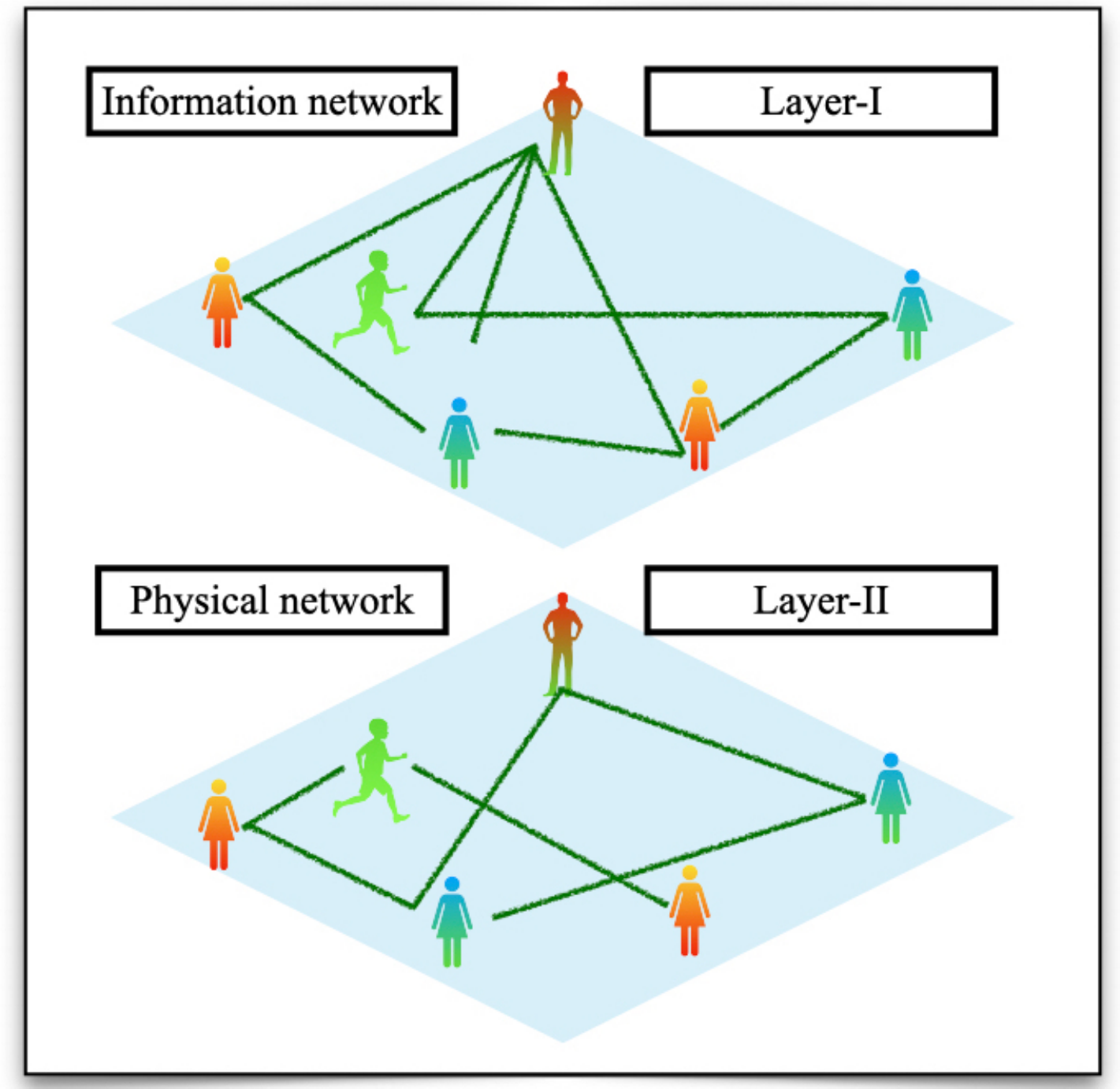}
\caption{Two layer multiplex network model. Human stickers represent the individuals in the community, and solid lines represent connections (virtual or physical) between the individuals.}
\label{FIG:1}
\end{figure}

In developing the coupled game-theoretic \& disease framework, we consider vaccination decision dynamics on a two-layer multiplex network (Fig. \ref{FIG:1} $\&$ \ref{FIG:2}), where \textit{Information layer} (layer-I) describes the social contacts among community members which is also adaptive in nature, and the \textit{Physical layer} (layer-II) depicts physical contacts among individuals that describe the spreading of infection in population. We have outlined below the detailed aspects of the model formulation, including assumptions, the decision-making process for vaccination, and the adaptive rules of social connections:\\

\noindent\textbf{Model assumptions}
\begin{enumerate}
    \item Networks in both layers have the same number of nodes, which is fixed throughout all analysis, but different numbers of connections.

    \item In the Information Network (layer-I), some old connections may be removed, and new connections may be added per day (described below).
\end{enumerate}

\subsection*{(a) Dynamics of disease transmission} 
\begin{figure}[h]
      \centering
      \includegraphics[scale=0.20]{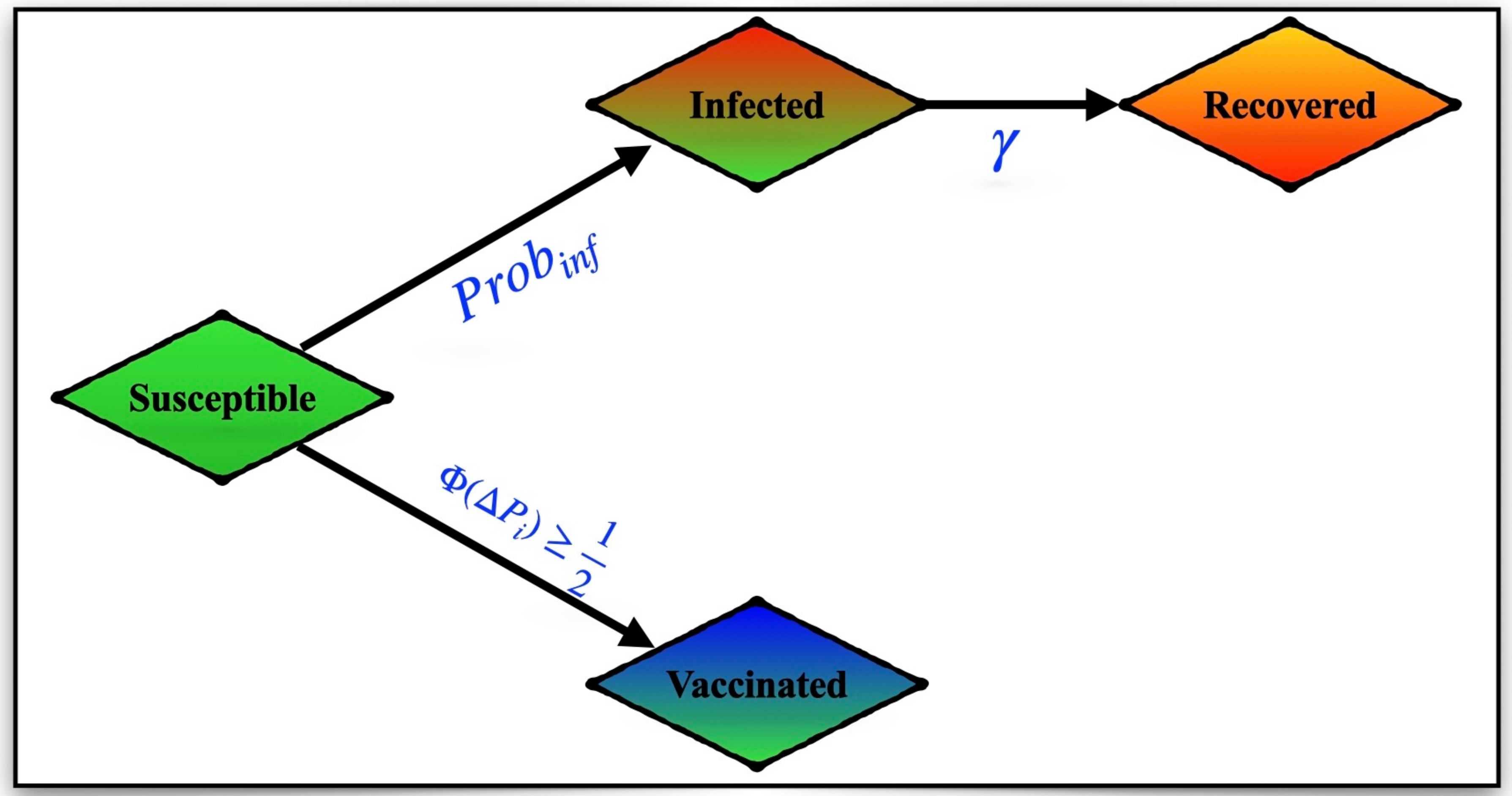}
      \caption{Schematic of the game-theoretic disease coupled model.}
      \label{FIG:2}
\end{figure}

We consider the standard SIR framework to model the disease spread on the network. We may assume influenza or COVID-19-like air-borne infections in this setting. The probability of acquiring infection for a $j^{th}$ susceptible node depends on its number of immediate infected neighbours ($N_{inf}^{j}$) and on the disease transmission rate ($\beta$). So, the transmission probability per day is defined by \cite{kumar2023nonlinear}:\\
     \begin{equation}
    Prob_{inf}\,=\,1-(1-\beta)^{N_{inf}^{j}} \label{eq:1}
     \end{equation}
The schematic representation of the disease dynamics in layer-II is shown in Figure \ref{FIG:2}.

\subsection*{(b) Dynamics of vaccination game} We assume there are two strategies: vaccinators and non-vaccinators. Individuals have the ability to switch between different strategies. The vaccination decision of an individual is a function of payoffs for both strategies, i.e., the perceived cost of infection and the perceived cost of vaccination, which influence the decision to be vaccinated. Let us assume that $f_j$ denotes the strategy of individual $j$ in the network, hence\\
\begin{equation}
f_j =
\begin{cases}
  1 & \text{if $j$ is vaccinators}\\
  0 & \text{if  $j$ is non-vaccinator}
\end{cases}\label{eq:2}
\end{equation}
Each $j^{th}$ individual has distinct levels of perceived vaccination risk ($C_{V,j}$) and perceived infection risk ($C_{I,j}$), which also evolves as described below. If $P_{NV,j}$ and $P_{V,j}$ are perceived payoff for a non-vaccination and vaccination of the $j^{th}$ individual, then\\
\begin{align}
\nonumber
    P(f_j = 1) &= P_{V,j} = -C_{V,j}\\
    P(f_j = 0) &= P_{NV,j} = -C_{I,j}*\theta_j \label{eq:3}
\end{align}
\noindent where $\theta_j$ is the perceived probability of infection that depends on the number of non-vaccinated neighbours:\\
\begin{equation*}
    \theta_j = \zeta[1-e^{-\varkappa N_{nv}^{j}}]
\end{equation*}
$N_{nv}^{j}$ denotes the number of non-vaccinated neighbours of the $j^{th}$ individual. Individuals may switch between strategies according to the Fermi-Dirac function \cite{bhattacharyya2019impact}:\\
\begin{equation*}
    \Phi(\Delta P_j) = \frac{1}{1+e^{-\xi*\Delta P_j}}
\end{equation*}
where $\Delta P_j$ is the payoff gain of the $j^{th}$ node given by $\Delta P_j$ $=$ $P_{V,j} - P_{NV,j}$. If
\begin{itemize} 
    \item $\Phi(\Delta P_j) \geq \frac{1}{2}$, the $j^{th}$ individual will choose to vaccinate with a probability $\Phi(\Delta P_j)$. 
    \item $\Phi(\Delta P_j) < \frac{1}{2}$, the $j^{th}$ individual will choose not to vaccinate. 
\end{itemize}
The perceived risk of infection $\theta_j$ and its effect on the probability of switching strategies significantly depend on the sensitivity of the focal individual to unvaccinated neighbours, the degree to which individuals respond to differences of payoff, and so on. So, the parameters $\varkappa$, $\xi$, and $\zeta$ are important in the dynamics of the vaccination game. Detailed description of the parameters $\varkappa$, $\xi$ and $\zeta$ are defined in the Table \ref{table1:values}. 

\subsection*{(c) Updating rules of individual risk perception}

We have categorized the vaccine risk perceptions in three levels or states: L $\{C_V \in [0, 0.25)\}$, M $\{C_V \in [0.25, 0.75)\}$, and H  $\{C_V \in [0.75, 1]\}$ (see Table \ref{table2:values}). Each node in the network receives a state at the beginning of the simulation (at time $t=0$) based on its own risk perception value. At each time step $t$, individuals update their own perceived risk to equal the average perceived risk of immediate neighbours with the most common risk category. For example, the state of a $j^{th}$ node is changed to category $H$ at a time $t$ if there are maximum numbers of immediate neighbours with category $H$, and its risk perception is updated as the average of $C_V$'s of all those neighbours with category $H$. However, in the case of multiple common categories, any one of them is chosen randomly.

\subsection*{(d) Evolution of social connections} 
In response to the updated vaccine risk perception at each time step, a dynamic rewiring process occurs in the social connections (layer-I) to adapt and optimize its response mechanisms. Mathematical expression for the rewiring process is as follows:\\

\noindent\textbf{(d.1) Removing existing connections:} For any two connected nodes, they may disconnect with some probability that increases with the increasing difference between their opinions. For example, the probability of disconnecting two connected nodes $i$ and $j$ per time-step is:\\
    \begin{equation*}
        Prob_D = \delta(1-e^{-\nu C_{V,ij}})
    \end{equation*}
    where $C_{V,ij}$ = $|C_{V,i}-C_{V,j}|$, and it is zero when individuals have the same $C_V$, and it increases as they become more different. $\delta$ and $\nu$ are some scaling parameters. \\
    
\noindent\textbf{(c.4) Introducing new connections in layer-I:} For any two disconnected nodes $i$ and $j$ in the layer-I, we introduce edge between them if $C_{V,ij}$ = $|C_{V,i}-C_{V,j}|$ $<$ $\epsilon$, where $\epsilon$ is a small positive real number. 
\begin{table}[h]
\caption{Baseline parameter values for simulation.}\label{table1:values}
\begin{tabular}{@{}lll@{}}
\toprule
 Parameter & Description & Value/Range \\
\midrule

$\beta$& disease transmission rate (per day) & 0.6  \cite{kumar2023nonlinear} \\
$\frac{1}{\gamma}$  & infectious period & 7 days \cite{kumar2023nonlinear} \\
$C_{V,j}$ & perceived vaccine risk of the $j^{th}$ individual & [0-1]\\
$C_{I,j}$ & perceived infection risk of the $j^{th}$ individual & [0-1]\\
$\varkappa$ & sensitivity of $j^{th}$ individual to non-vaccinated neighbours ($N_{nv}^{j}$) & 0.8, [0-1]\\
$\zeta$ & effective proportionality constant in perceived infection & 1, [0-1]\\
$\xi$ & degree of responsiveness to differences of payoff & 0.1, [0-1]\\
 $N_e$  & total number of nodes in the physical layer (layer-II) & 5000 \cite{kumar2023nonlinear}\\
$N_v$  & total number of nodes in the information layer (layer-I) & 5000 \cite{kumar2023nonlinear}\\
$\xi_e$  & total number of edges in the physical layer (layer-II) & 9997 \\
$\xi_v$  & total number of edges in the information layer (layer-I) & 149535 \\
$L(0)$ &  initial population having perceived vaccine risk $C_V$ $\in$ [0-0.25)  & 750\\
$M(0)$ &  initial population having perceived vaccine risk $C_V$ $\in$ [0.25-0.75)  & 3500\\
$H(0)$ &  initial population having perceived vaccine risk $C_V$ $\in$ [0.75-1]  & 750\\
\bottomrule
\end{tabular}
\end{table}

\section{Results}
\subsection{Computation of Epidemic threshold using the Microscopic Markov Chain approach}
\begin{figure}[h]
\centering
\includegraphics[scale=0.20]{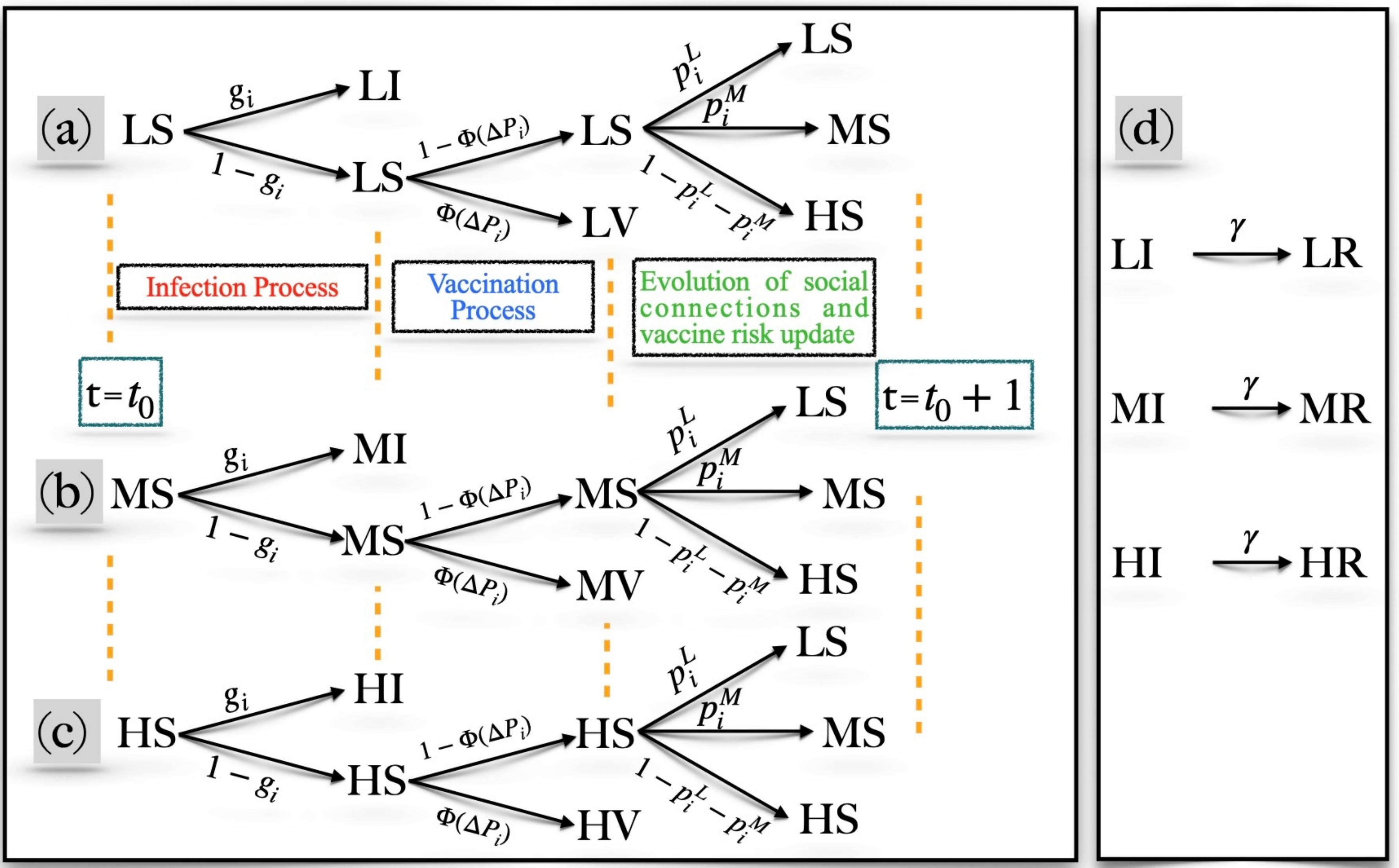}
\caption{(a) The state transition probability tree of a node in $LS$ state at time $t = t_0$. (b) The state transition probability tree of a node in $MS$ state at time $t = t_0$. (c) The state transition probability tree of a node in $HS$ state at time $t = t_0$. (d) The state transition probability tree of a node in $LI$, $MI$, and $HI$ state, respectively, is defined in Table \ref{table2:values}. Any other state that is the place to start has a transition probability tree that is a sub-tree of one of the trees that has been demonstrated.}
\label{FIG:21}
\end{figure}

We employed a microscopic Markov chain approach to derive the epidemic threshold in populations with varying risk perceptions. This approach, first introduced by Gómez et al. in 2010 \cite{gomez2010discrete}, analyzes the spread of contact-based epidemics by considering how the number of contacts evolves over time. In our analysis, we constructed transition trees to illustrate the progression of changes in node status over time (fig. \ref{FIG:21}).

For our analysis, we categorize the \(N\) nodes into nine distinct classes. At any given time \(t\), each node \(i\) can exist in one of the following nine states: $LV, MV, HV, LS, MS, HS, LI, MI$, and $HI$, with probabilities denoted by \(p_{i}^{LV}(t)\), \(p_{i}^{MV}(t)\), \(p_{i}^{HV}(t)\), \(p_{i}^{LS}(t)\), \(p_{i}^{MS}(t)\), \(p_{i}^{HS}(t)\), \(p_{i}^{LI}(t)\), \(p_{i}^{MI}(t)\), and \(p_{i}^{HI}(t)\) to traverse the path in transition trees. The detailed descriptions are provided in the appendix. Utilizing these probabilities and the transition tree, we calculate the epidemic threshold for infected individuals with various levels of perceived infection states such as Low ($LI$), Middle ($MI$), and Higher ($HI$). We derive the expression of the epidemic threshold as follows:\\
\begin{enumerate}
    \item \textbf{Epidemic threshold ($\beta^L$) for Low vaccine risk ($C_V \in [0, 0.25)$) community is:}
    \begin{align}
        \beta^{L} &= \frac{1}{\Delta_{max}(L)}
    \end{align}
    where $\Delta_{max}(L)$ is the largest eigenvalue of the matrix $L$ whose entries of $L$ matrix are:
    \begin{align}
        l_{ij} &= \{(1-\Phi(\Delta P_i))*p^{L}_i \{p_{i}^{LS}((1-p^{L}_i)(1-\Phi(\Delta P_i)) + 1 - {p_{i}^{LV}} - {p_{i}^{MV}} - {p_{i}^{HV}} \}\}b_{ij}\label{eq:M26}
    \end{align}
    \item \textbf{Epidemic threshold ($\beta^M$) for medium vaccine risk ($C_V \in [0.25, 0.75)$) community is:}
    \begin{align}
        \beta^{M} &= \frac{1}{\Delta_{max}(M)}
    \end{align}
    where entries of $M$ matrix are:
    \small{
    \begin{align}
        m_{ij} &= \{(1-\Phi(\Delta P_i))*p^{M}_i \{1+(1-p^{L}_i)(1-\Phi(\Delta P_i)\{1-({p_{i}^{MS}} + {p_{i}^{HS}} + {p_{i}^{LV}} + {p_{i}^{MV}} + {p_{i}^{HV}})\} \}\}b_{ij}
    \end{align}}
     \item \textbf{Epidemic threshold ($\beta^H$) for high vaccine risk ($C_V \in [0.75, 1]$) community is:}
    \begin{align}
        \beta^{H} &= \frac{1}{\Delta_{max}(H)}
    \end{align}
    where entries of $H$ matrix are:
    \small{
  \begin{align}
        h_{ij} &= \{(1-\Phi(\Delta P_i))*(1-p^{L}_i-p^{M}_i) \{p_{i}^{LS}((1-p^{L}_i)(1- \Phi(\Delta P_i)) + 1 - {p_{i}^{LV}} - {p_{i}^{MV}} - {p_{i}^{HV}} \}\}b_{ij}
    \end{align}}
    \end{enumerate}
The detailed derivation of all three epidemic thresholds is provided in the Appendix. Typically, in the absence of an information network (Layer I), the epidemic threshold is determined by the physical connections ($b_{ij}$), resulting in a single epidemic threshold for the community, as evident from the expressions for $l_{ij}$, $m_{ij}$, and $h_{ij}$ mentioned above. However, when vaccination behaviour and information dissemination within the population network are introduced, disproportionate vaccine risk perceptions emerge among individuals in the community, leading to changes in vaccination uptake. This alters the topology of the physical connections between susceptible and infected individuals, resulting in different patterns of epidemic thresholds. In the numerical simulations presented in the next section, we will demonstrate how these three epidemic thresholds vary across different parameter regimes.

\subsection{Numerical simulations}
We implement primarily a scale-free network in both layer-I and layer-II and use different values of power-law distribution parameters in the scale-free network, which determines the shape of the degree distribution of the network. We generate all of these networks using the igraph-python package \cite{csardi2006igraph,akhtar2014social}. Barabasi-Albert preferential attachment approach was implemented to generate scale-free networks \cite{barabasi2002evolution,newman2001clustering}. All of these networks have the same number of nodes: 5000. While the physical network (layer II) has 9997 edges, the virtual network (layer I) has a comparatively higher number of edges: 149535. Mean value of the degree distribution of physical network $\approx$ 4, whereas virtual network has larger mean $\approx$ 60. Degree distributions of all of these networks are shown in Fig. \textbf{S14 \& S15} in the supplementary information (\textbf{SI}). We also compute the Epidemic threshold (see figure \ref{FIG:data_epi_th}) for the social network within a selected village in southern India \cite{banerjee2013diffusion, sharma2019epidemic}. A detailed description of the dataset is in the \textbf{SI}. The supplementary materials provide a visual representation of the degree distribution of the social network (\textbf{S1, S2, \& S3}), total number of nodes, edges, and mean of the degree distribution of the dataset network are $1180$, $4626$, and $\approx$ 8 respectively. This analysis contributes to a broader understanding of epidemic dynamics within the context of social networks in contrast with scale-free networks. We used Python to execute all the simulations, and Matlab to plot all the simulated results.\\\\
\textbf{Computational simulation algorithm}: Here, we assume that a node in the physical network may be in one of four non-overlapping disease states: susceptible, infected, vaccinated, or recovered, and that nodes in the virtual layer can have one of three mutually exclusive perception-states: $L$, $M$, or $H$. The simulation algorithm's steps are as follows:
\begin{enumerate}
    \item Generate a multiplex network (with two layers) with $5000$ nodes. Layer-I ($G_1$) is the information network, and layer-II ($G_2$) is the physical network. 
    
    \item Initially, at t = 0, in absence of information network (layer I), we randomly have some initial infected nodes in the physical network (layer-II). Considering the baseline values of parameters as indicated in the Table \ref{table1:values}, We compute $Prob_{inf}$ and $\gamma$ as explained above, and execute simulations about 50 times. We repeat this exercise numerous times with different set of initial conditions and different parameter values within the range mentioned in the Table \ref{table1:values} to have final size of the epidemic approx 4848 in the community population (See fig \textbf{S19}). Taking this as standard disease transmission framework, we continue the simulation of entire model as follows. 
    
    \item We assign $C_{V,j}$ and $C_{I,j}$ randomly to all $j^{th}$ nodes in the network of layer-I. 
    
    \item We further calculate $\Delta P$ for each susceptible individual. Based on this, some susceptible individuals will choose vaccination, as explained above.
    
    \item We update the risk perception of each individual in layer-I as explained above.
    
    \item Remove old connections in layer-I and introduce new connections in layer-I as described above.
    
    \item Repeat steps 4-8 for the entire disease period.

    \item In presence of information dissemination and vaccinations the final size of epidemic is approx 3700. We consider this as standard dynamics to continue all our analyses.
    
\end{enumerate}
\subsubsection{Effect of variable risk perceptions and adaptive social connections on epidemic threshold}
\begin{figure}[h]
\centering
\includegraphics[scale=.35]{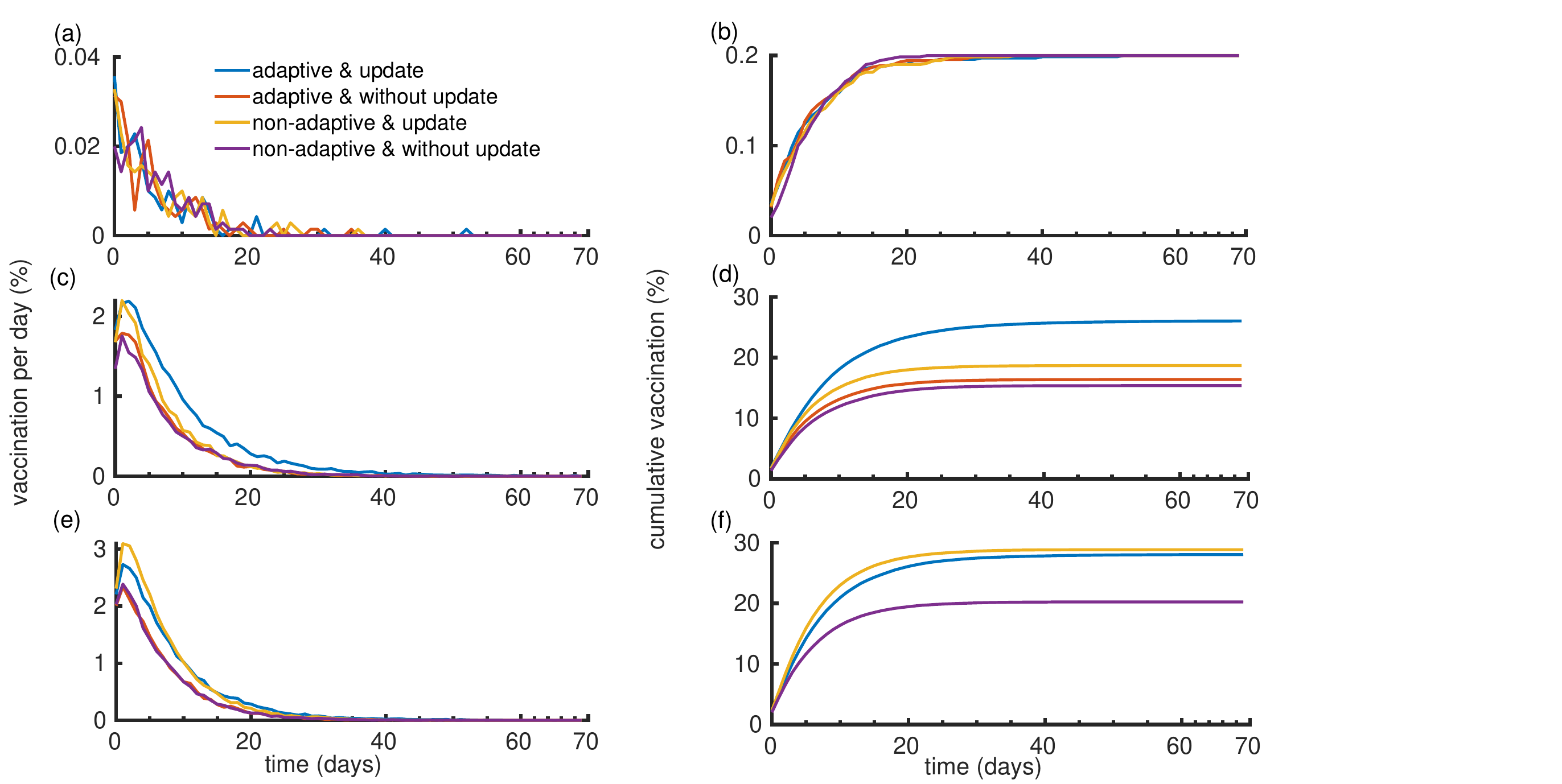}
\caption{Daily vaccine uptake and cumulative vaccination coverage respectively under (a-b) $\theta_j = 0$; (c-d) $\theta_j$ is defined as in section 2 ; and (e-f) $\theta_j = 1$. Here, `adaptive' indicated that removing and adding the connections in the information layer (layer-I) `non-adaptive' means static information network, and without update implies that each individual is not updating their vaccine risk perception as described in step 6 of the simulation algorithm. Each trajectory denotes an average of 50 simulations with the same parameter regime. Along with baseline values, other parameter values used for this simulation are $\delta = 0.8$, $\nu=1$ $\&$ $\epsilon=0.00022$.}
\label{FIG:mt1}
\end{figure}
We investigate how changes in perceived vaccine risk and the evolution of virtual connections influence people's choices to get a vaccine. Adaptivity in social connections and vaccination risk updates have an apparent impact on model dynamics, particularly on vaccine uptake (fig. \ref{FIG:mt1}).
A community's incentive to vaccinate diminishes when $\theta_j = 0$, as non-vaccinators receive no payoffs (fig. \ref{FIG:mt1} (a \& b)). But, if $\theta_j = 1$, the decision to vaccinate is mainly driven by risk of infection. Owing to the highest risk of infection, the adaptive nature of social connections  does not have much impact on choice of social infections (fig. \textbf{S16}). On a contrary, when $\theta_j$ is defined by above in section 2, there is moderate vaccine uptake ((fig. \ref{FIG:mt1} (c \& d))). However, the vaccine uptake is highest when the social connections is evolving. The risk of infection becomes variable and in presence of evolving social contacts, there is an initial increase non-vaccinator clusters in the population, which in turn, increase the risk perception of the infection, and hence the vaccination coverage. This result underscores that public health authority may conduct more vaccine campaign for proper risk communication about vaccine as well as risk of infection. \\

\begin{figure}[h]
      \centering      
      \includegraphics[scale=0.28]{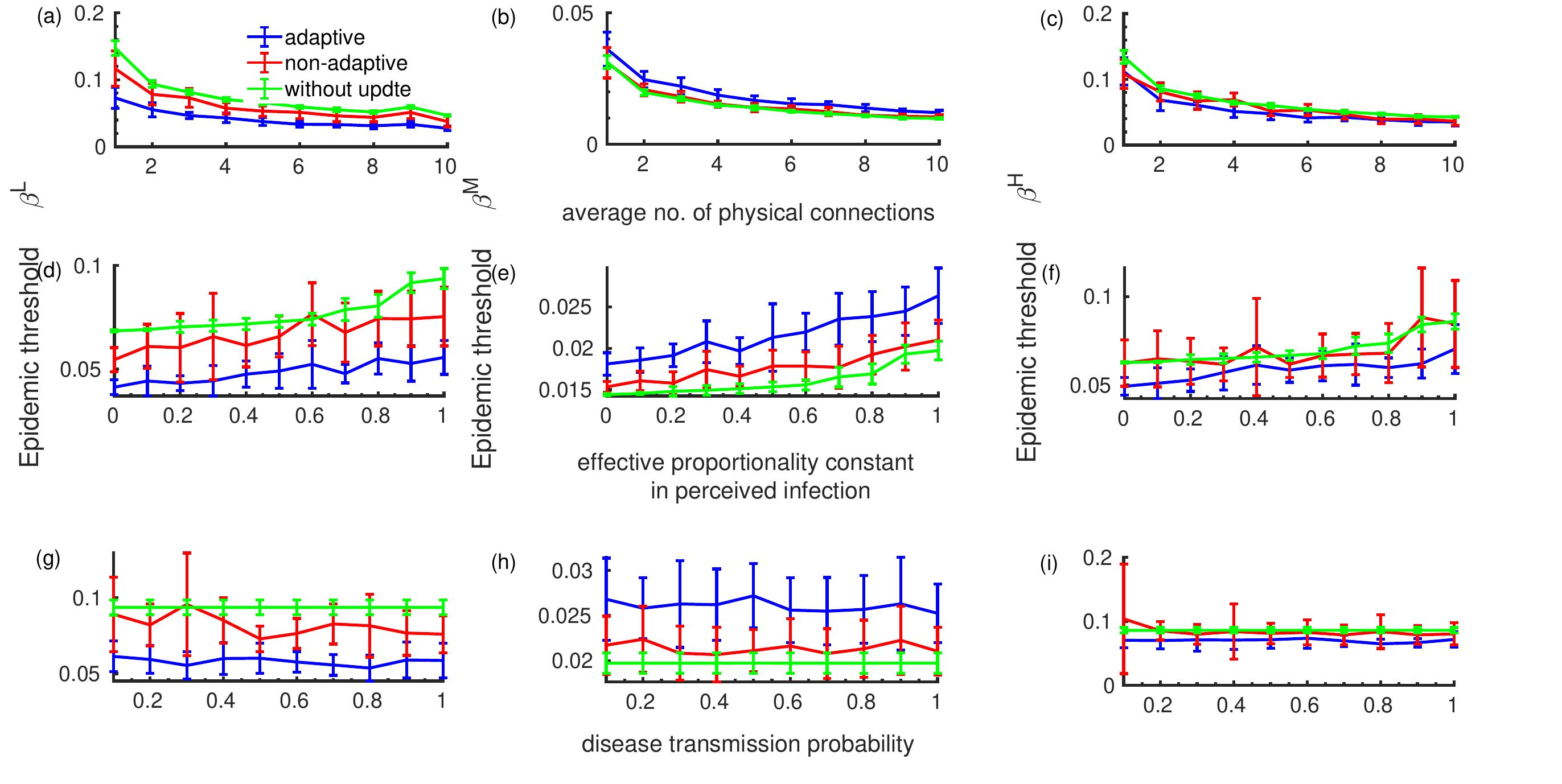}
      \caption{Plots of epidemic thresholds 
      $\beta^L$, $\beta^M$ and $\beta^H$ respectively for (a-b-c) different values of average physical connection in the Layer-II, (d -e-f) different proportions of the effective proportionality constant in perceived infection ($\zeta$), (g-h-i) different values of disease transmission rate ($\beta$). As indicated, each plot considers three different scenarios: (i) adaptive property of the virtual network (i.e., removing the old connection and introducing a new connection is conducted in Layer-I), (ii) non-adaptive property in the network i.e., the virtual network is static during the epidemic season. (iii) without update, i.e., none of the individuals in the virtual network updated their perceived vaccine risk during the entire time period of disease transmission. Plots in the first column, second column, and third column represent the epidemic threshold for low vaccine-risk people ($\beta^L$), medium vaccine-risk people ($\beta^M$), and high vaccine-risk people ($\beta^H$), respectively. Error bars display the standard deviations for 50 simulations. Along with baseline values, other parameter values used for this simulation are $\delta = 0.8$, $\nu=1$ $\&$ $\epsilon=0.00022$.}
      \label{FIG:epi_th}
\end{figure}
\begin{figure}[h]
      \centering
      \includegraphics[scale=0.28]{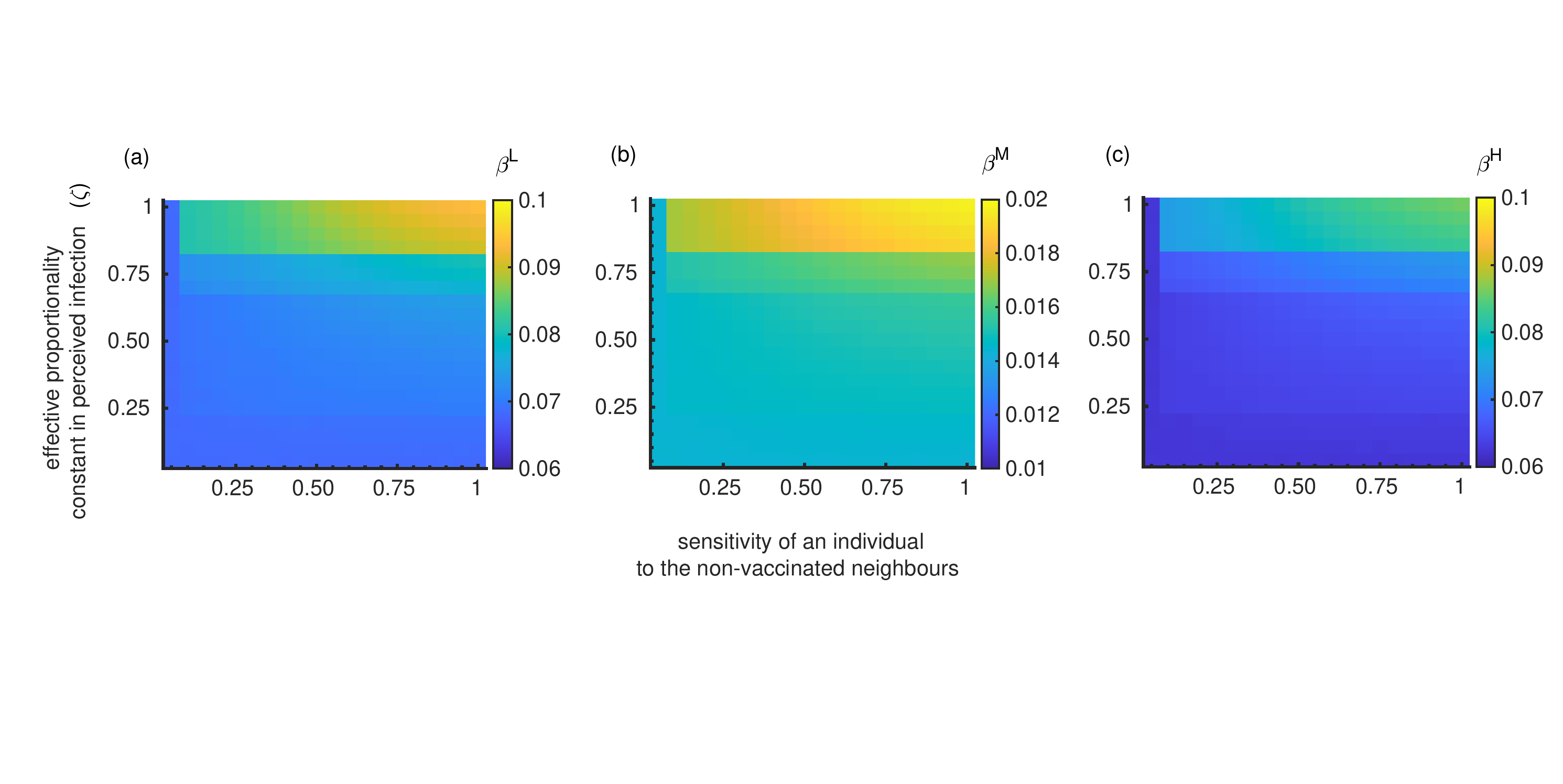}
      \vspace{-2.5cm}
      \caption{Two-parameter plot of the Epidemic threshold for (a) $LI$, (b) $MI$ and (c) $HI$ states as a function of effective proportionality constant in perceived infection ($\zeta$) and sensitivity of an individual to the non-vaccinated neighbours ($\varkappa$). Every point on the plot is the average of 25 simulations. Along with baseline values, other parameter values used for this simulation are $\delta = 0.8$, $\nu=1$ $\&$ $\epsilon=0.00022$.}
      \label{FIG:epi_surface}
\end{figure}

The epidemic threshold serves as a crucial indicator, revealing how quickly an outbreak spreads and establishes itself within a community's population. We conducted an in-depth analytical computation of the epidemic threshold, taking into account key model parameters, including the number of physical connections, the individual perception of vaccination risk ($\zeta$), and the transmission parameter ($\beta$). Figure \ref{FIG:epi_th} illustrates the qualitative changes in the epidemic thresholds under various values of these parameters. Several noteworthy dynamic changes were observed in this simulation.

Firstly, the epidemic threshold decreases as the average number of physical connections increases (Fig. \ref{FIG:epi_th}(a-c)), although the rate of decrease varies depending on the perceived risk of vaccination. This variation is evident from equations (4) to (9), which demonstrate that $\beta^i, ~~i=L,M,H$ are functions not only of physical connections $b_{ij}$ but also of the proportion of neighbours with low (L), medium (M), and high (H) perceived risks of vaccination within the community, which evolve over time. Consequently, the changes in the epidemic threshold differ disproportionately among these three groups. A similar phenomenon is observed when the effective proportionality constant ($\zeta$) is varied from 0-1 (Fig. \ref{FIG:epi_th}(a-c)). A higher $\zeta$ increases the risk of infection, leading to varying levels of vaccine uptake across different perceived risk groups, by changing their payoffs for vaccination. This in turn, raises the epidemic threshold. Interestingly, no significant change is observed when the transmission rate ($\beta$) is modified (Fig. \ref{FIG:epi_th}(g-f)), as adjusting this parameter does not alter the contact topology of individuals which is essential in computing the epidemic threshold (see equations (4-9)). 

Another interesting observation is that adaptive social connections cause the epidemic threshold to be consistently higher for individuals with a medium perceived risk of infection, but lower for those with either low or high perceived risk. This pattern emerges because evolving social connections lead to clusters of individuals with similar risk perceptions within the community. For groups with low or high perceived risk, this clustering tends to result in groups of non-vaccinators, whereas for the medium-risk group, it leads to higher vaccination rates. Of course, this observation can vary depending on the initial parameter values and the initial distribution of individuals among the three risk groups. In reality, most people in a community have a medium risk perception, so these findings may have significant implications for public health strategies aimed at controlling disease transmission in a population.          

Figure \ref{FIG:epi_surface}, demonstrates the effect of an individual's sensitivity to the neighbours and the effective proportionality constant on disease prevalence (fig. \ref{FIG:epi_surface}). We observed that as the $\varkappa$ $\&$ $\zeta$ increases, epidemic thresholds are increasing. Because higher values of these parameters reflect a higher vaccination rate, the disease burden is suppressed. However, the transition between $\beta^L$ $\&$ $\beta^H$ is higher than the $\beta^M$ (fig. \ref{FIG:epi_surface}). These variations in epidemic threshold are because individuals initially in $L$ and $H$ states tend to transition to the $M$ state, resulting in a larger proportion of the population in the latter.

\begin{figure}[h]
      \includegraphics[scale=0.25]{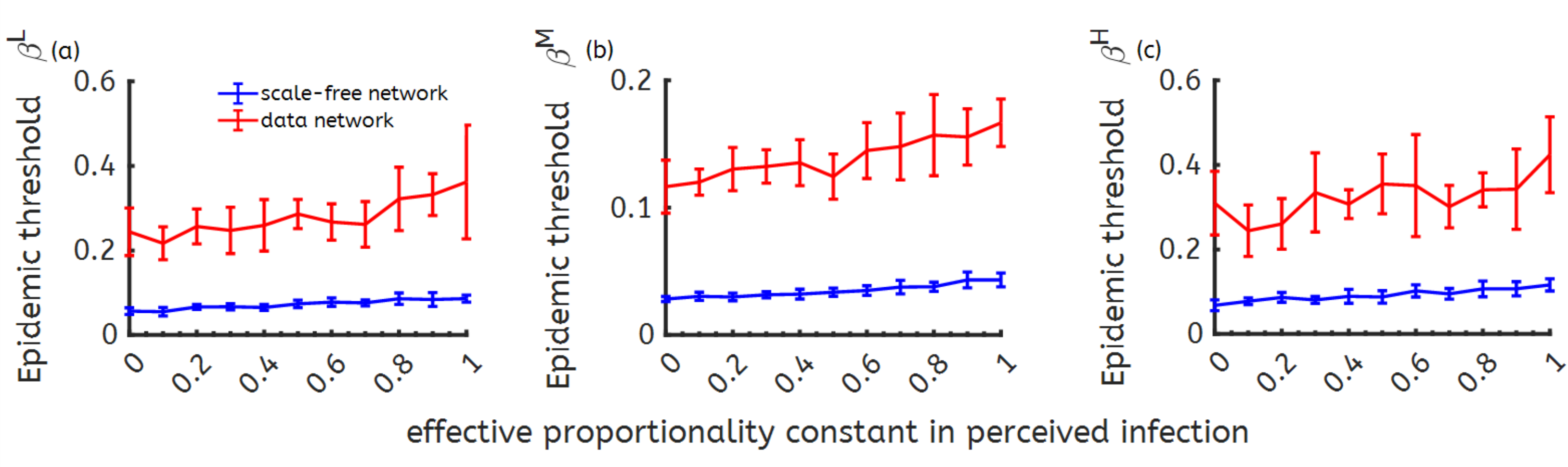}
      \caption{Plots of Epidemic threshold for (a) $LI$, (b) $MI$, and (c) $HI$ state for various values of the effective proportionality constant in perceived infection ($\zeta$). Here, the data network represents the real network of the southern villages in the physical layer, as explained above in the numerical section. The scale-free network uses the power-law exponent ($\alpha$) $=2.3$. The total number of nodes and edges for the scale-free network and data network are $1180$ and $4626$ respectively. Along with baseline values, other parameter values used for this simulation are $\delta = 0.8$, $\nu=1$ $\&$ $\epsilon=0.00022$. Error bars display the standard deviations for 50 simulations.}
      \label{FIG:data_epi_th}
\end{figure}

Simulations shown in Figures \ref{FIG:mt1}, \ref{FIG:epi_th} and \ref{FIG:epi_surface} were conducted using a scale-free contact network. Additionally, we calculated epidemic thresholds across empirically derived networks, as depicted in Figure \ref{FIG:data_epi_th}, for various $\zeta$ values. The epidemic threshold tends to be higher in empirical networks than in scale-free networks (fig. \ref{FIG:data_epi_th}), due to the lower incidence of hubs, despite a high mean degree (fig. \textbf{S1}). This reduces the speed of disease spread, delaying the onset of an outbreak. 

Taken together, these simulation results may have policy insights for controlling disease. policymakers may acquire valuable insights into the impact of different vaccination rates and non-vaccination behaviours on epidemic thresholds. To effectively mitigate the impact of epidemics, it is essential to ascertain which demographic subgroups are most vulnerable to disease transmission and to identify crucial thresholds for disease transmission. In addition, our research highlights the significance of community-wide vaccination campaigns to increase population immunity rates and decrease infectious disease burdens.

\subsubsection{Impact of the individual's sensitivity to non-vaccinated neighbours ($\varkappa$) and effective proportionality constant in perceived infection ($\zeta$) on perceived probability of infection}
\begin{figure}[h]
\centering
\includegraphics[scale=0.30]{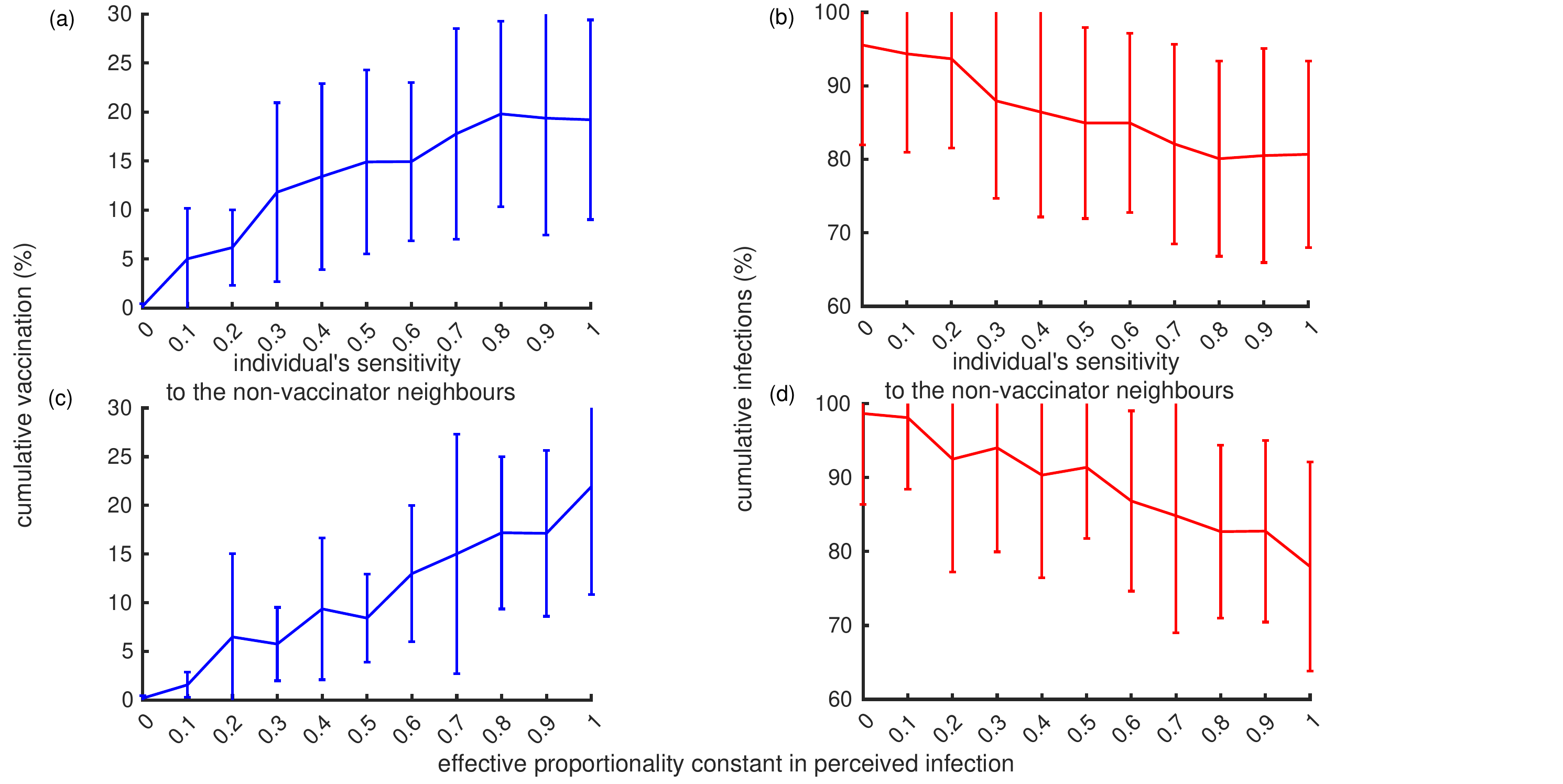}
\caption{Figure depicts (a) cumulative vaccination ($\%$); (b) cumulative infection ($\%$) as a function of the sensitivity of an individual to non-vaccinated neighbours ($\varkappa$); (c-d) represent same variables as a function of effective proportionality constant in perceived infection ($\zeta$). Error bars depicted on the plot represent the standard deviation calculated from 50 simulations for each data point. Along with baseline values, other parameter values used for this simulation are $\delta = 0.8$, $\nu=1$ $\&$ $\epsilon=0.00022$.}
\label{FIG:combine_c_z}
\end{figure}
Sensitivity to non-vaccinated neighbours can vary among individuals in a population, potentially influencing their perceptions of personal risk. The likelihood of contracting an infection from non-vaccinated neighbours is influenced by the strength of their connections. For instance, close connections, such as those within families or schools, can lead to faster and more widespread transmission compared to interactions in workplaces or public spaces \cite{mogi2022influence,leung2021transmissibility}. The rate of transmission can also differ depending on the type of pathogen involved. Additionally, individuals' awareness and attitudes toward the disease, often influenced by socio-economic factors \cite{cerda2021hesitation,al2021factors}, play a significant role in shaping their perceptions of vaccination. The parameters $\varkappa$ and $\zeta$ serve as proxies for such phenomena observed in reality. 

We have simulated our model for different values of $\varkappa$ and $\zeta$ to better capture these effects (fig. \ref{FIG:combine_c_z}). Figure \ref{FIG:combine_c_z} (a $\&$ b) depicts our model's simulation for different levels of the individual’s sensitivity to non-vaccinated neighbours that contribute to the perceived possibility of infection. A discernible trend is evident, demonstrating an escalation in cumulative vaccination concurrent with an increase in parameter $\varkappa$, leading to a corresponding reduction in cumulative infections within the community. We systematically simulate our model for various values of $\varkappa$ (refer to fig. \ref{FIG:combine_c_z} (a $\&$ b)). Specifically, for $\varkappa = 0.1$, the increase in cumulative vaccination is $5.1 \%$, whereas at $\varkappa = 0.5$ and $0.8$, the cumulative vaccinated populations are increased to $14.90 \%$ and $19.80 \%$, respectively (see fig. \ref{FIG:combine_c_z} (a)). Simultaneously, the cumulative burden of infections in the community is diminishing (see fig. \ref{FIG:combine_c_z} (b)). As the value of $\varkappa$ increases, individuals will get more sensitive to have non-vaccinated neighbours, which in turn, increase the $\theta_j$, which  leads to a increase in pay-off for vaccination. Thus, more people likely to get vaccines for higher values of $\varkappa$ (see fig. \ref{FIG:combine_c_z} (a)), and eventually decrease in cumulative disease burden (fig. \ref{FIG:combine_c_z} (b)).

We observe a similar impact of the effective proportionality constant ($\zeta$) in perceived infection. Figure \ref{FIG:combine_c_z} (c $\&$ d) demonstrates the 
qualitative changes in the cumulative vaccination and infection due to change in $\zeta$. The cumulative vaccination increases and cumulative disease burden decreases in a relatively more linear fashion when the $\zeta$ increases. Thus, the observed relationship between connection strength, disease transmissibility, and an individual's attitude toward a pathogen can provide important insights that might inform targeted measures during outbreak. Incorporating these results into public health initiatives can pave the way for a more sophisticated and focused strategy that will help communities prevent and control infectious diseases.\\

\subsubsection{Effect of power law exponent on the disease incidence}
\begin{figure}[h]
      \centering
      \includegraphics[scale=0.32]{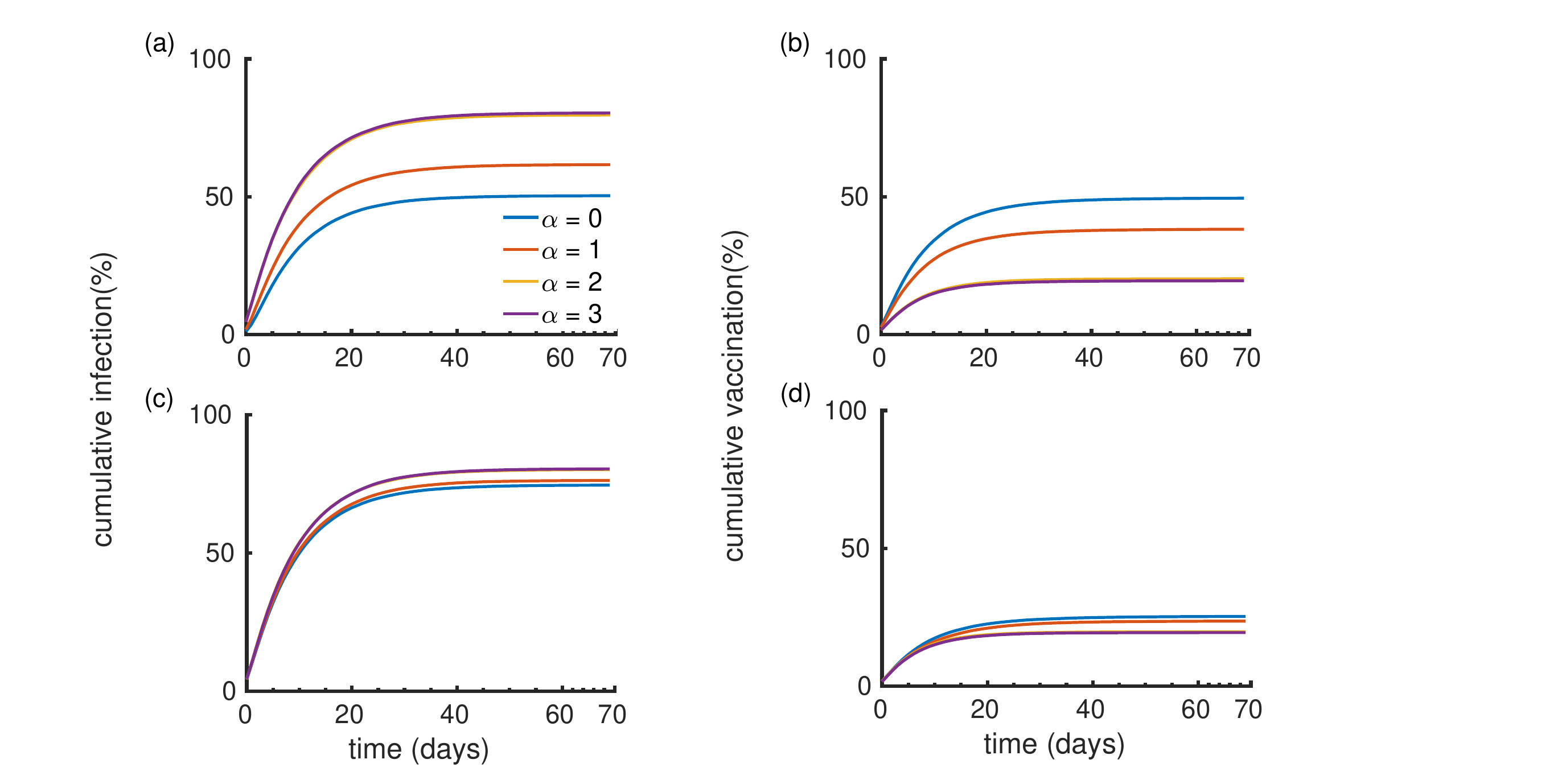}
      \caption{(a) Figure depicts (a) cumulative infected ($\%$); (b) cumulative vaccinated; for different values of power-law exponent ($\alpha$) in the physical network with $\alpha = 2.3$ in the virtual network (c) cumulative infected ($\%$); (d) cumulative vaccinated; for different values of power-law exponent ($\alpha$) in the virtual network with $\alpha = 2.3$ in the physical network. Each point on the plots is the average of 50 iterations. Along with baseline values, other parameter values used for this simulation are $\delta = 0.8$, $\nu=1$ $\&$ $\epsilon=0.00022$}
      \label{FIG:alha}
\end{figure}
Physical connections are critical in navigating the super spread of disease, as they represent key points where targeted interventions, such as vaccinations, can significantly curb transmission \cite{liu2016identify}. A critical aspect of network metrics is the power-law exponent ($\alpha$), which determines the distribution of connections within the network and has a profound influence on model behaviour \cite{ndeffo2012impact,danon2012social}. A smaller $\alpha$ value indicates a more evenly distributed network of social interactions across the population \cite{mossa2002truncation}. In contrast, a higher $\alpha$ value suggests a highly heterogeneous network, where a small number of individuals (hubs) possess a disproportionately large number of connections \cite{broido2019scale,meng2023scale}.
Figure \ref{FIG:alha} demonstrates how altering the power-law exponent affects the cumulative dynamics of vaccination and infection. We observe a low cumulative infection rate for $\alpha = 0$ in the physical network (fig. \ref{FIG:alha} (a)), reflecting a more homogeneous distribution of contacts with fewer disease transmissions (fig. \textbf{S6 \& S7}). As a result, the daily incidence of infections decreases. Fewer infections lead to a higher chance for individuals to revise their perceptions of the risk of vaccinations (fig. \textbf{S18} (b)), increasing the cumulative vaccination (fig. \ref{FIG:alha} (b)). Since there is more diversified physical contact (hubs) (fig. \textbf{S8 \& S9}) for $\alpha = 2$ or $3$, the $Prob_{inf}$ increases drastically (fig. \ref{FIG:alha} (a)). That will cause an increase in the infected population (fig. \textbf{S18} (a)). Modifying the power-law exponent ($\alpha$) within the virtual network yields minimal alterations in disease prevalence (fig. \ref{FIG:alha} (c,d)). Notably, as $\alpha$ increases, the majority of nodes manifest homogeneous degree distributions (fig. \textbf{S12 \& S13}), minimizing the network's influence on risk updating dynamics. As $\alpha = 0$, virtual nodes exhibit heterogeneous degrees (fig. \textbf{S10}), resulting in slightly elevated vaccine rates compared to scenarios with higher $\alpha$ values in virtual connections (fig. \ref{FIG:alha} (d)).

Effect of the various power-law exponents on vaccination and infection rates has critical implications for public health strategies. By understanding how changes in network topology, especially in terms of $\alpha$ values, affect vaccination and infection dynamics, policymakers and healthcare practitioners can gain valuable insights for designing targeted interventions. Incorporating these insights into decision-making can significantly improve the effectiveness of vaccination campaigns within communities.\\

\section{Discussion and conclusion}
Amidst the widespread growth of online social media platforms, the exchange of information and behaviours related to vaccination have a substantial impact on vaccine uptake, which indirectly influences the control of infectious disease outbreaks \cite{yin2022impact, kumar2023nonlinear, guarino2022automatic}. People now routinely use social media for advice and information on various health-related topics, including vaccination decisions, epidemic awareness, and the ethical implications of antibiotic use \cite{bae2021accounting, scanfeld2010dissemination, krishnan2021research, au2021role}. On the other hand, people tend to interact more with others who have similar thoughts in the context of disease perceptions, often forming connections through mutual acquaintances. Therefore, as social networks continue to evolve, it creates  disproportionate risk perceptions among individuals in society, which in turn, influences their decisions to support the control strategies planned by public health authorities to control the outbreak in the community.\\

Several key insights from our investigation bear significance in the realms of population decision-making in vaccination, network evolution, and disease transmission in the community. We observed a significant impact of adaptive social connections on three epidemic thresholds, influenced by variations in individuals' perceived vaccine risk. Specifically, the epidemic threshold increased for individuals with a moderate level of perceived vaccine risk. This finding suggests that public health authorities should intensify vaccine campaigns and improve risk communication to encourage individuals with higher perceived risk to shift towards a moderate level of perceived risk. By doing so, the overall epidemic threshold of the population can be raised, thereby reducing the likelihood of an outbreak. The physical connections within the network structure also emerged as a critical factor influencing epidemic threshold. For example, higher power-law exponents impede vaccine uptake as they accelerated infection spread among individuals, thereby infecting individuals before they had the chance to get vaccinated. Our study also emphasizes that individual sensitivity to vaccinated neighbours play roles in enhancing payoff differences and ultimately promoting increased vaccine uptake in the population. These findings offer valuable insights for public health authorities, emphasizing the need for proper risk communication about vaccines and the severity of infections to increase the awareness in the community. Such strategic measures hold potential for curbing infections in the community.\\

The present research provides valuable insights into the connections between perceived vaccine risk, vaccine awareness, the structure of individuals' social networks, and the resulting impact on vaccine uptake. However, it is important to acknowledge the potential limitations of our model and the possibility for improvement by incorporating more realistic scenarios. In real-world situations, especially when a vaccine is newly introduced, the risks associated with it are often unknown. This uncertainty can lead to the spread of rumors on social media about adverse effects, such as paralysis, lung congestion, or even death. These adverse events can disproportionately influence perceived vaccine risk across an entire community \cite{liu2022coevolution,mushtaq2022review, bhattacharyya2019impact}. Similar considerations are relevant for entirely new pathogens like COVID-19. Incorporating these factors into our model could alter the qualitative aspects of our findings, potentially aiding in the development of more effective infection control policies.

Another aspect that warrants attention is the adaptability of physical networks. Individuals typically avoid physical contact with sick relatives and friends and adopt social isolation measures during a pandemic. As a result, physical networks evolve over time. Our model could be further refined by incorporating these dynamic changes, and this presents scope for future research. \\

Our study provides a potential framework for developing more effective and tailored strategies to control vaccine-preventable diseases by analyzing the interplay between vaccine choices, awareness dissemination, and network structures. In an increasingly interconnected world, enhanced disease control plans and the advancement of public health programs can be achieved by leveraging social media platforms to convey accurate information about vaccines and infections.\\

\section{CRediT authorship contribution statement}
\textbf{Viney Kumar:} Coding, Methodology, Formal analysis, Writing – original draft. \textbf{Chris T Bauch:} Conceptualization, Methodology, Formal analysis, Writing – review \& editing. \textbf{Samit Bhattacharyya:} Conceptualization, Methodology, Formal analysis, Writing – review \& editing, Supervision.
\\\\

\section{Data availability}
Code supporting the findings of this study can be accessed via the following link: \url{https://github.com/vineykumar2801/Epidemicthreshold-adaptivenetwork.git}

\section{Declaration of competing interest}
The authors declare that they have no known competing financial interests or personal relationships that could have
appeared to influence the work reported in this paper.\\

\section{Acknowledgments}
First author is thankful to the Council of Scientific \& Industrial Research (CSIR), India, Senior Research Fellowship scheme (FILE NUMBER 09/1128(12030)/2021-EMR-I) for financial support. First and Last author acknowledge the Department of Mathematics, Shiv Nadar Institution of Eminence (SNIoE) for DST-FIST-funded computational lab support.\\

\appendix
\section{Appendix}
\subsection{Computation of Epidemic threshold using the Microscopic Markov Chain approach}
    We explore the vaccine risk-exchanging-epidemic coupling dynamics in the multiplex networks using a Microscopic Markov Chain approach. We represent the probability that node $i$ is in $X$ state at time t as $p_{i}^{X}(t)$, where $X\in (L,M,H,S,I,R,V,LV,MV,HV,LS,MS,HS,LI, MI, HI)$, (see table \ref{table2:values}) where the set $(L,M,H)$ are states of nodes in the information layer denoting the order of perceptions in vaccine risk, the set $(S,I,R,V)$ are the states of the nodes in the physical layer, and the set $(LV,MV,HV,LS,MS,HS,LI,MI,HI)$ are the states of nodes in the vaccine risk-exchanging-epidemic coupled dynamics. Here, it's worth noting that the order of individuals' perceptions regarding vaccination influences their decision-making process, which subsequently impacts the transmission dynamics within the physical network. Accordingly, we calculate three distinct epidemic thresholds corresponding to different perception states among individuals: (i) the epidemic threshold for lower-order perceptions denoted as $\beta^L$, (ii) the epidemic threshold for middle-order perceptions denoted as $\beta^M$, and (iii) the epidemic threshold for higher-order perceptions denoted as $\beta^H$. These thresholds are computed individually from the equations of the coupled dynamics (for more clarifications, see Figure \ref{FIG:21}): 

\begin{table}[h]
\caption{States of all the nodes used in the computation of epidemic threshold}\label{table2:values}
\begin{tabular}{@{}ll@{}}
\toprule
State of the nodes & Description \\
\midrule
$L$  & node having perceived vaccine risk $C_V$ $\in$ [0-0.25) \\
$M$  & node having perceived vaccine risk $C_V$ $\in$ [0.25-0.75) \\
$H$ & node having perceived vaccine risk $C_V$ $\in$ [0.75-1] \\
$S$ & susceptible node \\
$I$ & infected node\\
$R$ & recovered node \\
$V$  & vaccinated node  \\
$LS$  & susceptible node in the layer - II having $L$ state in layer-I\\
$MS$  & susceptible node in the layer - II having $M$ state in layer-I \\
$HS$  & susceptible node in the layer - II having $H$ state in layer-I \\
$LV$  & vaccinated node in the layer - II having $L$ state in layer-I \\
$MV$  & vaccinated node in the layer - II having $M$ state in layer-I \\
$HV$  & vaccinated node in the layer - II having $H$ state in layer-I \\
$LI$  & infected node having $L$ state in layer-I\\
$MI$  & infected node having $M$ state in layer-I \\
$HI$  & infected node having $H$ state in layer-I \\
$LR$  & recovered node having $L$ state in layer-I\\
$MR$  & recovered node having $M$ state in layer-I \\
$HR$  & recovered node having $H$ state in layer-I \\
\bottomrule
\end{tabular}
\end{table}
\begin{table}[h]
\caption{Symbols used in the computation of epidemic threshold}\label{table3:values}
\begin{tabular}{@{}ll@{}}
\toprule
Symbol & Description \\
\midrule
$a_{ij}$  & elements in the adjacency matrix of information layer \\
$b_{ij}$  & elements in the adjacency matrix of physical layer \\
$g_{i}(t)$ & probability of node $i$ being infected by its neighbour at time $t$ \\
$p_{i}^{X}(t)$ & probability of node $i$ being in state $X$ at time $t$ \\
$p_{i}^{X}$ & probability of node $i$ being in state $X$ in the stationary state \\
$p_j^{L}(t)$ & probability of $j^{th}$ node having neighbours of $L$ state in the layer-I at time t\\
$p_j^{M}(t)$ & probability of $j^{th}$ node having neighbours of $M$ state in the layer-I at time t\\
$p_j^{H}(t)$ & probability of $j^{th}$ node having neighbours of $H$ state in the layer-I at time t\\
\bottomrule
\end{tabular}
\end{table}

    \begin{align}
        {p_{i}^{LS}}(t+1) &= p_{i}^{LS}(t)*(1-g_{i}(t))*(1-\Phi(\Delta P_i)(t))*p^{L}_i(t)+p_{i}^{LS}(t)*(1-g_{i}(t))*(1-\Phi(\Delta P_i)(t))\nonumber\\&{}*p^{M}_i(t)*(1-g_{i}(t))*(1-\Phi(\Delta P_i)(t))*p^{L}_i(t)+p_{i}^{MS}(t)*(1-g_{i}(t))*(1-\Phi(\Delta P_i)(t))*p^{L}_i(t)\nonumber\\&{} + p_{i}^{LS}(t)*(1-g_{i}(t))*(1-\Phi(\Delta P_i)(t))*(1-p^{L}_i(t)-p^{M}_i(t))*(1-g_{i}(t))*(1-\Phi(\Delta P_i)(t))\nonumber\\&{}*p^{L}_i(t)+ p_{i}^{HS}(t)*(1-g_{i}(t))*(1-\Phi(\Delta P_i)(t))*p^{L}_i(t) \label{eq:M2}\\
        {p_{i}^{MS}}(t+1) &= p_{i}^{LS}(t)*(1-g_{i}(t))*(1-\Phi(\Delta P_i)(t))*p^{M}_i(t)*(1-g_{i}(t))*(1-\Phi(\Delta P_i)(t))*p^{M}_i(t) +p_{i}^{MS}(t)\nonumber\\&{}*(1-g_{i}(t))*(1-\Phi(\Delta P_i)(t))*p^{M}_i(t)+p_{i}^{LS}(t)*(1-g_{i}(t))*(1-\Phi(\Delta P_i)(t))*\nonumber\\&{}(1-p^{L}_i(t)-p^{M}_i(t))*(1-g_{i}(t))*(1-\Phi(\Delta P_i)(t))*p^{M}_i(t)+p_{i}^{HS}(t)*(1-g_{i}(t))*\nonumber\\&{}(1-\Phi(\Delta P_i)(t))*p^{M}_i(t) \label{eq:M3}\\
       {p_{i}^{HS}}(t+1) &= p_{i}^{LS}(t)*(1-g_{i}(t))*(1-\Phi(\Delta P_i)(t))*p^{M}_i(t)*(1-g_{i}(t))*(1-\Phi(\Delta P_i)(t))*\nonumber\\&{}(1-p^{L}_i(t)-p^{M}_i(t))+p_{i}^{MS}(t)*(1-g_{i}(t))*(1-\Phi(\Delta P_i)(t))*(1-p^{L}_i(t)-p^{M}_i(t))+\nonumber\\&{}p_{i}^{LS}(t)*(1-g_{i}(t))*(1-\Phi(\Delta P_i)(t))*(1-p^{L}_i(t)-p^{M}_i(t))*(1-g_{i}(t))*(1-\Phi\nonumber\\&{}(\Delta P_i)(t))*(1-p^{L}_i(t)-p^{M}_i(t))+p_{i}^{HS}(t)*(1-g_{i}(t))*(1-\Phi(\Delta P_i)(t))*(1-p^{L}_i(t)-p^{M}_i(t)) \label{eq:M4}\\
        {p_{i}^{LV}}(t+1) &= p_{i}^{LS}(t)*(1-g_{i}(t))*\Phi(\Delta P_i)(t) \label{eq:M5}\\
        {p_{i}^{MV}}(t+1) &= p_{i}^{LS}(t)*(1-g_{i}(t))*(1-\Phi(\Delta P_i)(t))*p^{M}_i(t)*(1-g_{i}(t))*\Phi(\Delta P_i)(t)+p_{i}^{MS}(t)\nonumber\\&{}*(1-g_{i}(t))*\Phi(\Delta P_i)(t) \label{eq:M6}\\
        {p_{i}^{HV}}(t+1) &= p_{i}^{LS}(t)*(1-g_{i}(t))*(1-\Phi(\Delta P_i)(t))*(1-p^{L}_i(t)-p^{M}_i(t))*(1-g_{i}(t))*\Phi(\Delta P_i)(t)\nonumber\\&{}+p_{i}^{HS}(t)*(1-g_{i}(t))*\Phi(\Delta P_i)(t) \label{eq:M7}\\
        {p_{i}^{LI}}(t+1) &= p_{i}^{LS}(t)*g_{i}(t)\label{eq:M8}\\
        {p_{i}^{MI}}(t+1) &= p_{i}^{LS}(t)*(1-g_{i}(t))*(1-\Phi(\Delta P_i)(t))*p^{M}_i(t)*g_{i}(t)+p_{i}^{MS}(t)g_{i}(t)\label{eq:M9}\\
        {p_{i}^{HI}}(t+1) &= p_{i}^{LS}(t)*(1-g_{i}(t))*(1-\Phi(\Delta P_i)(t))*(1-p^{L}_i(t)-p^{M}_i(t))*g_{i}(t)+p_{i}^{HS}(t)*g_{i}(t)\label{eq:M10}\\
        {p_{i}^{LR}}(t+1) &= p_{i}^{LI}(t)+p_{i}^{LR}(t)\label{eq:M11}\\
         {p_{i}^{MR}}(t+1) &= p_{i}^{MI}(t)+p_{i}^{MR}(t)\label{eq:M12}\\
          {p_{i}^{HR}}(t+1) &= p_{i}^{HI}(t)+p_{i}^{HR}(t)\label{eq:M13}\\
        \nonumber
    \end{align}
    In the final state, i.e., $t\to\infty$ in Eqs. (\ref{eq:M2})-(\ref{eq:M10}), there is 
    \begin{align}
    \nonumber
        {p_{i}^{X}}(t+1) &= {p_{i}^{X}}(t) = {p_{i}^{X}}
    \end{align}
    Near the epidemic threshold, the proportion of infected nodes to the total population is low enough; hence, we can assume that the probability of a node becoming infected is \ref{table3:values}
    \begin{align}
    \nonumber
    1-g_{i}(t) &= 1-\beta\sum\limits_{j}b_{ij}\epsilon_{j}=1-\eta_{i}
    \end{align}
    Since the number of infected nodes may be disregarded close to the threshold, ${p_{i}^{LR}}$$\to0$, ${p_{i}^{MR}}$$\to0$ and ${p_{i}^{HR}}$$\to0$. Eqs. (\ref{eq:M2})-(\ref{eq:M7}) can be modified as
    \begin{align}
      {p_{i}^{LS}}(t+1) &= p_{i}^{LS}(t)*(1-\Phi(\Delta P_i)(t))*p^{L}_i(t) + p_{i}^{LS}(t)*(1-\Phi(\Delta P_i)(t))*p^{M}_i(t)*(1-\Phi(\Delta P_i)(t))\nonumber\\&{}*p^{L}_i(t)+p_{i}^{MS}(t)*(1-\Phi(\Delta P_i)(t))*p^{L}_i(t)+p_{i}^{LS}(t)*(1-\Phi(\Delta P_i)(t))*(1-p^{L}_i(t)-p^{M}_i(t))\nonumber\\&{}*(1-\Phi(\Delta P_i)(t))*p^{L}_i(t)+ p_{i}^{HS}(t)*(1-\Phi(\Delta P_i)(t))*p^{L}_i(t) \label{eq:M14}\\
        {p_{i}^{MS}}(t+1) &= p_{i}^{LS}(t)*(1-\Phi(\Delta P_i)(t))*p^{M}_i(t)*(1-\Phi(\Delta P_i)(t))*p^{M}_i(t) +p_{i}^{MS}(t)*(1-\Phi(\Delta P_i)(t))\nonumber\\&{}*p^{M}_i(t)+p_{i}^{LS}(t)*(1-\Phi(\Delta P_i)(t))*(1-p^{L}_i(t)-p^{M}_i(t))*(1-\Phi(\Delta P_i)(t))*p^{M}_i(t)+p_{i}^{HS}(t)\nonumber\\&{}*(1-\Phi(\Delta P_i)(t))*p^{M}_i(t) \label{eq:M15}\\
       {p_{i}^{HS}}(t+1) &= p_{i}^{LS}(t)*(1-\Phi(\Delta P_i)(t))*p^{M}_i(t)*(1-\Phi(\Delta P_i)(t))*(1-p^{L}_i(t)-p^{M}_i(t))+p_{i}^{MS}(t)*\nonumber\\&{}(1-\Phi(\Delta P_i)(t))*(1-p^{L}_i(t)-p^{M}_i(t))+p_{i}^{LS}(t)*(1-\Phi(\Delta P_i)(t))*(1-p^{L}_i(t)-p^{M}_i(t))*\nonumber\\&{}(1-\Phi(\Delta P_i)(t))*(1-p^{L}_i(t)-p^{M}_i(t))+p_{i}^{HS}(t)*(1-\Phi(\Delta P_i)(t))*(1-p^{L}_i(t)-p^{M}_i(t)) \label{eq:M16}\\
        {p_{i}^{LV}}(t+1) &= p_{i}^{LS}(t)*\Phi(\Delta P_i)(t) \label{eq:M17}\\
        {p_{i}^{MV}}(t+1) &= p_{i}^{LS}(t)*(1-\Phi(\Delta P_i)(t))*p^{M}_i(t)*\Phi(\Delta P_i)(t)+p_{i}^{MS}(t)*\Phi(\Delta P_i)(t) \label{eq:M18}\\
        {p_{i}^{HV}}(t+1) &= p_{i}^{LS}(t)*(1-\Phi(\Delta P_i)(t))*(1-p^{L}_i(t)-p^{M}_i(t))*\Phi(\Delta P_i)(t)+p_{i}^{HS}(t)*\Phi(\Delta P_i)(t) \label{eq:M19}\\
        \nonumber
    \end{align}
    \subsubsection{Epidemic threshold ($\beta^{L}$) for low vaccination risk ($C_V \in [0,0.25))$ community}
    To compute the epidemic threshold ($\beta^L$) for $LI$ state, we derive the corresponding equation (\ref{eq:M8}) in the system's stationary state as follows:
    \begin{align}
        \epsilon_{i} &= p_{i}^{LS}*\eta_{i} \label{eq:M20}\\
        \nonumber
    \end{align}
    Now, substituting Eq. (\ref{eq:M14}) in Eq. (\ref{eq:M20}), we get
    \begin{align*}
        \epsilon_{i} &= [p_{i}^{LS}*(1-\Phi(\Delta P_i))*p^{L}_i + p_{i}^{LS}*(1-\Phi(\Delta P_i))*p^{M}_i*(1-\Phi(\Delta P_i))*p^{L}_i+p_{i}^{MS}\nonumber\\&{}*(1-\Phi(\Delta P_i))*p^{L}_i+p_{i}^{LS}*(1-\Phi(\Delta P_i))*(1-p^{L}_i-p^{M}_i)*(1-\Phi(\Delta P_i))*p^{L}_i+ \nonumber\\&{}p_{i}^{HS}*(1-\Phi(\Delta P_i))*p^{L}_i]*\eta_i
    \end{align*}
    \begin{align}
        \epsilon_{i} &= [(1-\Phi(\Delta P_i))*p^{L}_i \{p_{i}^{LS}(1+(1-p^{L}_i)(1-\Phi(\Delta P_i))+p_{i}^{MS}+p_{i}^{HS}\}]*\eta_i \label{eq:M21}\\
        \nonumber
    \end{align}
    Near the epidemic threshold, we get
    \begin{align}
        {p_{i}^{LS}} + {p_{i}^{MS}} + {p_{i}^{HS}} + {p_{i}^{LV}} + {p_{i}^{MV}} + {p_{i}^{HV}} &= 1\label{eq:M22}
    \end{align}
    \begin{align}
        {p_{i}^{MS}} + {p_{i}^{HS}} &= 1 -{p_{i}^{LS}} - ({p_{i}^{LV}} + {p_{i}^{MV}} + {p_{i}^{HV}} )\label{eq:M23}
    \end{align}
    Putting, Eq. (\ref{eq:M23}) into Eq. (\ref{eq:M21}), we get
    \begin{align*}
    \nonumber
        \epsilon_{i} &= [(1-\Phi(\Delta P_i))*p^{L}_i \{p_{i}^{LS}(1+(1-p^{L}_i)(1-\Phi(\Delta P_i))+ 1 -{p_{i}^{LS}} - ({p_{i}^{LV}} + {p_{i}^{MV}} + {p_{i}^{HV}} )\}]*\eta_i\\
        =& [(1-\Phi(\Delta P_i))*p^{L}_i \{p_{i}^{LS}((1-p^{L}_i)(1-\Phi(\Delta P_i))+ 1 - {p_{i}^{LV}} - {p_{i}^{MV}} - {p_{i}^{HV}} \}]*\eta_i
    \end{align*}
    \begin{align}
        &= [(1-\Phi(\Delta P_i))*p^{L}_i \{p_{i}^{LS}((1-p^{L}_i)(1-\Phi(\Delta P_i))+ 1 - {p_{i}^{LV}} - {p_{i}^{MV}} - {p_{i}^{HV}} \}]*\beta\sum\limits_{j}b_{ij}\epsilon_{j}\label{eq:M24}
    \end{align}
    Eq. (\ref{eq:M24}) can be expressed as a matrix form as follows:
    \begin{align}
        \sum\limits_{j}[\{(1-\Phi(\Delta P_i))*p^{L}_i \{p_{i}^{LS}((1-p^{L}_i)(1-\Phi(\Delta P_i)) + 1 - {p_{i}^{LV}} - {p_{i}^{MV}} - {p_{i}^{HV}} \}\}b_{ij} - \frac{1}{\beta}\sigma_{ij}]*\epsilon_{j} &= 0\label{eq:M25}
    \end{align}
    $\sigma_{ij}$ represents the elements of identity matrix. The lowest value that satisfies Eq. (\ref{eq:M24}) is referred to as the epidemic threshold ($\beta^{L}$). We can simplify the equation to the eigenvalue problem for matrix $L$ in order to get the epidemic threshold as a self-consistent equation. The elements of the matrix $L$ are:
    \begin{align}
        l_{ij} &= \{(1-\Phi(\Delta P_i))*p^{L}_i \{p_{i}^{LS}((1-p^{L}_i)(1-\Phi(\Delta P_i)) + 1 - {p_{i}^{LV}} - {p_{i}^{MV}} - {p_{i}^{HV}} \}\}b_{ij}\label{eq:M26}
    \end{align}
    Let $\Delta_{max}($L$)$ be the largest eigenvalue of the matrix $L$, the epidemic threshold ($\beta^{L}$) will be:
    \begin{align}
        \beta^{L} &= \frac{1}{\Delta_{max}(L)}\label{eq:M27}
    \end{align}
    It shows that the epidemic threshold ($\beta^{L}$) depends on number of physical connection, probability $p_{i}^{LS}$, $p_{i}^{LV}$, $p_{i}^{MV}$, $p_{i}^{HV}$, $\Phi(\Delta P_i)$ and $p^{L}_i$ of $i^{th}$ node.

    \subsubsection{Epidemic threshold ($\beta^{M}$) for medium vaccination risk ($C_V \in [0.25,0.75))$ community}
    For finding the epidemic threshold for the $MI$ state we will express the equation (\ref{eq:M9}) as follows:
    \begin{align}
        \epsilon_{i} &= [ p_{i}^{LS}*(1-\Phi(\Delta P_i))*p^{M}_i+p_{i}^{MS} ] * \eta_i \label{eq:M28}\\
        \nonumber
    \end{align}
    Now, substituting Eq. (\ref{eq:M15}) in Eq. (\ref{eq:M28}), we get
    \begin{align*}
        \epsilon_{i} &= [p_{i}^{LS}*(1-\Phi(\Delta P_i))*p^{M}_i + p_{i}^{LS}*(1-\Phi(\Delta P_i))*p^{M}_i*(1-\Phi(\Delta P_i))*p^{M}_i+p_{i}^{MS}\nonumber\\&{}*(1-\Phi(\Delta P_i))*p^{M}_i+p_{i}^{LS}*(1-\Phi(\Delta P_i))*(1-p^{L}_i-p^{M}_i)*(1-\Phi(\Delta P_i))*p^{M}_i+ \nonumber\\&{}p_{i}^{HS}*(1-\Phi(\Delta P_i))*p^{M}_i]*\eta_i
    \end{align*}
    \begin{align}
        \epsilon_{i} &= [(1-\Phi(\Delta P_i))*p^{M}_i \{p_{i}^{LS}(1+(1-p^{L}_i)(1-\Phi(\Delta P_i))+p_{i}^{MS}+p_{i}^{HS}\}]*\eta_i \label{eq:M29}\\
        \nonumber
    \end{align}
    Near the epidemic threshold, we get
    \begin{align}
        {p_{i}^{LS}} + {p_{i}^{MS}} + {p_{i}^{HS}} + {p_{i}^{LV}} + {p_{i}^{MV}} + {p_{i}^{HV}} &= 1\label{eq:M30}
    \end{align}
    \begin{align}
        {p_{i}^{LS}} &= 1 -{p_{i}^{MS}} -  {p_{i}^{HS}} - ({p_{i}^{LV}} + {p_{i}^{MV}} + {p_{i}^{HV}} )\label{eq:M31}
    \end{align}
    Putting, Eq. (\ref{eq:M31}) into Eq. (\ref{eq:M29}), we get
    \begin{align*}
    \nonumber
        \epsilon_{i} &= [(1-\Phi(\Delta P_i))*p^{M}_i \{(1 -{p_{i}^{MS}} -  {p_{i}^{HS}} - ({p_{i}^{LV}} + {p_{i}^{MV}} + {p_{i}^{HV}} ))(1+(1-p^{L}_i)(1-\Phi(\Delta P_i))+ {p_{i}^{MS}} +  {p_{i}^{HS}}\}]*\eta_i\\
        =& [(1-\Phi(\Delta P_i))*p^{M}_i \{1+(1-p^{L}_i)(1-\Phi(\Delta P_i)\{1-{p_{i}^{MS}} - {p_{i}^{HS}} - {p_{i}^{LV}} - {p_{i}^{MV}} - {p_{i}^{HV}}\}\}]*\eta_i
    \end{align*}
    \begin{align}
        &= [(1-\Phi(\Delta P_i))*p^{M}_i \{1+(1-p^{L}_i)(1-\Phi(\Delta P_i)\{1-({p_{i}^{MS}} + {p_{i}^{HS}} + {p_{i}^{LV}} + {p_{i}^{MV}} + {p_{i}^{HV}})\}\}]*\beta\sum\limits_{j}b_{ij}\epsilon_{j}\label{eq:M32}
    \end{align}
    Eq. (\ref{eq:M32}) can be expressed as a matrix form as follows:
    \begin{align}
        \sum\limits_{j}[\{(1-\Phi(\Delta P_i))*p^{M}_i \{1+(1-p^{L}_i)(1-\Phi(\Delta P_i)\{1-({p_{i}^{MS}} + {p_{i}^{HS}} + {p_{i}^{LV}} + {p_{i}^{MV}} + {p_{i}^{HV}})\} \}\}b_{ij} - \frac{1}{\beta}\sigma_{ij}]*\epsilon_{j} &= 0\label{eq:M33}
    \end{align}
    $\sigma_{ij}$ represents the elements of identity matrix. The lowest value that satisfies Eq. (\ref{eq:M32}) is referred to as the epidemic threshold ($\beta^{M}$). We can simplify the equation to the eigenvalue problem for matrix $M$ in order to get the epidemic threshold as a self-consistent equation. The elements of the matrix $M$ are:
    \begin{align}
        m_{ij} &= \{(1-\Phi(\Delta P_i))*p^{M}_i \{1+(1-p^{L}_i)(1-\Phi(\Delta P_i)\{1-({p_{i}^{MS}} + {p_{i}^{HS}} + {p_{i}^{LV}} + {p_{i}^{MV}} + {p_{i}^{HV}})\} \}\}b_{ij}\label{eq:M34}
    \end{align}
    Let $\Delta_{max}($M$)$ be the largest eigenvalue of the matrix $M$, the epidemic threshold ($\beta^{M}$) will be:
    \begin{align}
        \beta^{M} &= \frac{1}{\Delta_{max}(M)}\label{eq:M35}
    \end{align}
    It shows that the epidemic threshold ($\beta^{M}$) depends on number of physical connection, probability $p_{i}^{MS}$, $p_{i}^{HS}$, $p_{i}^{LV}$, $p_{i}^{MV}$, $p_{i}^{HV}$, $\Phi(\Delta P_i)$, $p^{L}_i$ and $p^{M}_i$ of $i^{th}$ node.

    \subsubsection{Epidemic threshold ($\beta^{H}$) for higher vaccination risk ($C_V \in [0.75,1])$ community}
    For finding the epidemic threshold for the $HI$ state we will express the equation (\ref{eq:M10}) as follows:
    \begin{align}
        \epsilon_{i} &= [ p_{i}^{LS}*(1-\Phi(\Delta P_i))*p^{M}_i+p_{i}^{MS} ] * \eta_i \label{eq:M36}\\
        \nonumber
    \end{align}
    Now, substituting Eq. (\ref{eq:M16}) in Eq. (\ref{eq:M36}), we get
    \begin{align*}
        \epsilon_{i} &= [p_{i}^{LS}*(1-\Phi(\Delta P_i))*(1-p^{L}_i-p^{M}_i) + p_{i}^{LS}*(1 -\Phi(\Delta P_i))*p^{M}_i*(1-\Phi(\Delta P_i))*\nonumber\\&{}(1-p^{L}_i-p^{M}_i)+p_{i}^{MS}*(1-\Phi(\Delta P_i))*(1-p^{L}_i-p^{M}_i)+p_{i}^{LS}*(1-\Phi  (\Delta P_i))*(1-p^{L}_i\nonumber\\&{}-p^{M}_i)*(1-\Phi(\Delta P_i))*(1-p^{L}_i-p^{M}_i)+p_{i}^{HS}*(1-\Phi(\Delta P_i))*(1-p^{L}_i-p^{M}_i)]*\eta_i
    \end{align*}
    \begin{align}
        \epsilon_{i} &= [(1-\Phi(\Delta P_i))*(1-p^{L}_i-p^{M}_i) \{p_{i}^{LS}(1+(1-p^{L}_i)(1 -\Phi(\Delta P_i))+p_{i}^{MS}+p_{i}^{HS}\}]*\eta_i \label{eq:M37}\\
        \nonumber
    \end{align}
    Near the epidemic threshold, we get
    \begin{align}
        {p_{i}^{LS}} + {p_{i}^{MS}} + {p_{i}^{HS}} + {p_{i}^{LV}} + {p_{i}^{MV}} + {p_{i}^{HV}} &= 1\label{eq:M38}
    \end{align}
    \begin{align}
        {p_{i}^{MS}} + {p_{i}^{HS}} &= 1 -{p_{i}^{LS}} - ({p_{i}^{LV}} + {p_{i}^{MV}} + {p_{i}^{HV}} )\label{eq:M39}
    \end{align}
    Putting, Eq. (\ref{eq:M39}) into Eq. (\ref{eq:M37}), we get
    \begin{align*}
    \nonumber
        \epsilon_{i} &= [(1-\Phi(\Delta P_i))*(1-p^{L}_i-p^{M}_i) \{p_{i}^{LS}(1+(1-p^{L}_i)(1 -\Phi(\Delta P_i))+ 1 -{p_{i}^{LS}} - ({p_{i}^{LV}} + {p_{i}^{MV}} + {p_{i}^{HV}} )\}]*\eta_i\\
        =& [(1-\Phi(\Delta P_i))*(1-p^{L}_i-p^{M}_i) \{p_{i}^{LS}((1-p^{L}_i)(1-\Phi(\Delta P_i ))+ 1 - {p_{i}^{LV}} - {p_{i}^{MV}} - {p_{i}^{HV}} \}]*\eta_i
    \end{align*}
    \begin{align}
        &= [(1-\Phi(\Delta P_i))*(1-p^{L}_i-p^{M}_i) \{p_{i}^{LS}((1-p^{L}_i)(1-\Phi(\Delta P_i ))+ 1 - {p_{i}^{LV}} - {p_{i}^{MV}} - {p_{i}^{HV}} \}]*\beta\sum\limits_{j}b_{ij}\epsilon_{j}\label{eq:M40}
    \end{align}
    Eq. (\ref{eq:M40}) can be expressed as a matrix form as follows:
    \begin{align}
        \sum\limits_{j}[\{(1-\Phi(\Delta P_i))*(1-p^{L}_i-p^{M}_i) \{p_{i}^{LS}((1-p^{L}_i)(1-\Phi( \Delta P_i )) + 1 - {p_{i}^{LV}} - {p_{i}^{MV}} - {p_{i}^{HV}} \}\}b_{ij} - \frac{1}{\beta}\sigma_{ij}]*\epsilon_{j} &= 0\label{eq:M41}
    \end{align}
    $\sigma_{ij}$ represents the elements of identity matrix. The lowest value that satisfies Eq. (\ref{eq:M40}) is referred to as the epidemic threshold ($\beta^{H}$). We can simplify the equation to the eigenvalue problem for matrix $H$ in order to get the epidemic threshold as a self-consistent equation. The elements of the matrix $H$ are:
    \begin{align}
        h_{ij} &= \{(1-\Phi(\Delta P_i))*(1-p^{L}_i-p^{M}_i) \{p_{i}^{LS}((1-p^{L}_i)(1- \Phi(\Delta P_i)) + 1 - {p_{i}^{LV}} - {p_{i}^{MV}} - {p_{i}^{HV}} \}\}b_{ij}\label{eq:M42}
    \end{align}
    Let $\Delta_{max}($H$)$ be the largest eigenvalue of the matrix $H$, the epidemic threshold ($\beta^{H}$) will be:
    \begin{align}
        \beta^{H} &= \frac{1}{\Delta_{max}(H)}\label{eq:M43}
    \end{align}
    It shows that the epidemic threshold ($\beta^{H}$) depends on number of physical connection, probability $p_{i}^{LS}$, $p_{i}^{LV}$, $p_{i}^{MV}$, $p_{i}^{HV}$, $\Phi(\Delta P_i)$, $p^{L}_i$ and $p^{M}_i$ of $i^{th}$ node.

\bibliography{ref}

\end{document}